\documentclass[preprint2]{emulateapj} 
\usepackage{xspace}
\usepackage{color}
\usepackage{natbib}
\usepackage{appendix}



\long\def\comment#1{}

\def\W2{{\cal W}}

\def\bi{\begin{itemize}}
\def\ei{\end{itemize}}
\def\be{\begin{equation}}
\def\ee{\end{equation}}
\def\bea{\begin{eqnarray}}
\def\eea{\end{eqnarray}}

\def\cmm2{{\,\rm cm^{-2}}}
\def\cm2{{\,{\rm cm}^2}}
\def\cmm3{{\,{\rm cm}^{-3}}}
\def\gcmm3{{\,{\rm g\,cm^{-3}}}}

\def\fun#1#2{\lower3.6pt\vbox{\baselineskip0pt\lineskip.9pt
  \ialign{$\mathsurround=0pt#1\hfil##\hfil$\crcr#2\crcr\sim\crcr}}}

\def\W{{\cal W}}

\def\Planck{\textit{Planck}\xspace}

\def\muK2{$\mu$K$^2$}
\newcommand{\Ground}{Ground\xspace}
\newcommand{\fig}{Figure\xspace}
\newcommand{\tab}{Table\xspace}
\newcommand{\eq}{Eq.\xspace}
\xspaceaddexceptions{+}

\begin{document}

\lefthead{}
\righthead{}


\title{Modeling Extragalactic Foregrounds and Secondaries for Unbiased Estimation of
  Cosmological Parameters from Primary CMB Anisotropy}

\author{ M. Millea\altaffilmark{1}, O. Dor\'e\altaffilmark{2,3},
  J. Dudley\altaffilmark{4}, G. Holder\altaffilmark{4},  L. Knox\altaffilmark{1},
 L. Shaw\altaffilmark{5}, Y.-S. Song\altaffilmark{6}, O. Zahn\altaffilmark{7}}
\altaffiltext{1}{Department of Physics, University of California, One
 Shields Avenue, Davis, CA 95616}
\altaffiltext{2}{Jet Propulsion Laboratory, California Institute of Technology, Pasadena, CA 91109}
\altaffiltext{3}{California Institute of Technology, Pasadena, CA 91125}
\altaffiltext{4} {Department of Physics, McGill University, 3600 Rue University, Montreal, Quebec H3A 2T8, Canada}
\altaffiltext{5}{Department of Physics, Yale University, P.O. Box 208210, New Haven, CT 06520-8120}
\altaffiltext{6}{Korea Institute for Advanced Study, Dongdaemun-gu, Seoul 130-722, Korea}
\altaffiltext{7}{Berkeley Center for Cosmological Physics, Department of Physics, University of California, and Lawrence Berkeley National Labs, Berkeley, CA 94720}

\vspace{1cm}

\begin{abstract}
  Using the latest physical modeling and constrained by the most recent
  data, we develop a phenomenological parameterized model of the contributions to
  intensity and polarization maps at millimeter wavelengths from
  external galaxies and Sunyaev-Zeldovich effects. We find such
  modeling to be necessary for estimation of cosmological parameters
  from \Planck data. For example, ignoring the clustering of the
  infrared background would result in a bias in $n_s$ of 7$\sigma$.
  We show that the simultaneous marginalization over a full
  foreground model can eliminate such biases, while
  increasing the statistical uncertainty in cosmological parameters by
  less than 20\%.  The small increases in uncertainty can be
  significantly reduced with the inclusion of higher-resolution ground-based data.
 
  The multi-frequency analysis we employ involves modeling 46 total
  power spectra and marginalization over 17 foreground parameters.  We
  show that we can also reduce the data to a best estimate of the CMB
  power spectra, and just two principal components (with constrained
  amplitudes) describing residual foreground contamination.
\end{abstract}

\keywords{cosmological parameters --- distance scale --- 
large-scale structure of universe}

\bigskip\bigskip


\vspace{1cm}

\section{Introduction}

The cosmic microwave background is arguably the most powerful probe of
the parameters and viability of cosmological models.  With temperature
measurements on angular scales larger than a third of a degree already
at the cosmic variance limit \citep{komatsu10}, further progress now
depends on improvements at smaller angular scales as well as improved
measurements of the polarization at all angular scales.  Advances in
temperature anisotropy measurements come most recently from the
Atacama Cosmology Telescope (ACT; \citet{dunkley10}), soon to be
followed by new results from the South Pole Telescope (SPT;
Keisler et al. 2011).  By the end of 2012 we expect a dramatic
improvement from \Planck, which is likely to leave very little
room for further improvement in the measurement of the primary CMB
temperature power spectrum.  These improved measurements will
translate into much tighter constraints on cosmological models
\citep[e.g.][]{planckbluebook}.

At angular scales smaller than a tenth of a degree, 
extragalactic foregrounds\footnote{We will
henceforth refer to both extragalactic foreground contaminants and 
secondary anisotropies as just 
``foregrounds" since they cannot be modeled from first principles like the primary CMB.}  
become important for three reasons: 1) the CMB power spectrum is
dropping in amplitude, 2) cosmic variance is smaller and 3)
foregrounds are growing in amplitude.  At sufficiently
small angular scales, foregrounds become the
dominant signal at all CMB frequencies.  Furthermore, unlike
galactic foregrounds, they are statistically isotropic and thus cannot
be avoided by masking regions of higher contamination.  Their modeling
is an unavoidable necessity.

In this paper we present a parameterized, physically-motivated,
phenomenological model for the extragalactic foregrounds and consider
it in the context of extracting cosmological parameters from the
primary CMB anisotropy. We demonstrate that for an analysis of \Planck
data, such modeling is necessary to avoid significant biases in
cosmological parameter estimates, but that marginalization over even
a very rich foreground model is essentially ``for free"; 
the foregrounds are sufficiently orthogonal to the primary CMB that 
the statical errors on cosmological parameters are degraded by at most 20\%
for $n_s$ and less than 10\% for other parameters. With the addition
of higher resolution ground-based data or non-CMB \Planck bands to clean 
the foregrounds, the degradation is reduced to a few percent for all parameters.

The importance of extragalactic foregrounds for CMB analysis has been
recognized for a long time \citep{Tegmark:1995pn, Bouchet:1999gq,
  knox99, Tegmark:1999ke, Leach:2008fi,smica, dunkley10}.  Potential
biases from extragalactic contaminants have been pointed out
previously by \citet{knox98,santos03,zahn05,serra08} and
\citet{taburet09}. Distinguishing our work is the simultaneous consideration
of all foreground components necessary for an analysis of \Planck data,
and physical modeling of these components
informed from recent measurements beyond the damping tail
by SPT \citep{hall10, vieira10, shirokoff10} and ACT \citep{dunkley10}.
In this paper, we will consider the foreground power contributions
from shot noise due to radio galaxies and dusty star forming galaxies
(DSFGs), the clustering of the DSFGs, the thermal and kinetic
Sunyaev-Zeldovich effects (tSZ and kSZ), and correlation between the
tSZ and DSFG components. We now turn to summarizing recent
developments in both modeling and measurements of these extragalactic
foregrounds.

Our understanding of the power spectrum due to DSFGs at frequencies
relevant for CMB analysis has been rapidly improving.
We demonstrate here that for analysis of \Planck data, the effects of
DSFG clustering are the most important of the foregrounds to
model.  Although it is the most important effect, it has been almost
entirely ignored by previous cosmological parameter error forecasting
work.  To date, the only papers to consider the impact of DSFG
clustering on cosmological parameter estimates are \citet{dunkley10}
and \citet{serra08}.

DSFG clustering power was first detected at CMB frequencies by the SPT
\citep{hall10}, with subsequent confirmation and improved constraints
from ACT \citep{dunkley10} and SPT \citep{shirokoff10}.  The
recent suite of early \Planck papers \citep[in
particular]{planckcib} have also provided significant constraints on both the
amplitude and shape of the clustering power.  The \Planck
measurements rule out many otherwise viable models which generally
predict higher power (on the scales relevant for analysis of the
primary CMB power spectrum) than observed.

Radio galaxy source counts from high-resolution ground-based data are
particularly useful for \Planck since they are sensitive to the decade
in brightness below \Planck's flux cut.  The radio sources in this
brightness range create the dominant source of shot noise power in
most of the \Planck frequencies which contain significant CMB
information.  SPT measurements of point source populations
\citep{vieira10} have offered valuable information about the
amplitude of Poisson power, as well as the coherence of these
shot-noise fluctuations from frequency to frequency.

Recent data, as well as recent theoretical developments, inform our
modeling of the power spectrum of the tSZ effect---a spectral
distortion that arises due to inverse Compton scattering of CMB
photons off the hot electrons in groups and clusters.  The magnitude
of the tSZ signal is proportional to the thermal pressure of the
intra-cluster medium (ICM) integrated along the line of sight.  Upper
limits on the amplitude of the tSZ power (set by \citet{lueker10},
confirmed by \citet{dunkley10} and further tightened by
\citet{shirokoff10}) were found to be surprisingly low compared to
predictions from halo model calculations \citep{komatsu02} and
non-radiative hydrodynamical simulations \citep{white02}.  Recent work
has demonstrated that the inclusion of a significant non-thermal
contribution to the total gas pressure in groups and clusters in
analytic models can significantly reduce the predicted amplitude of
the tSZ power spectrum \citep{shaw10, trac10}. Non-thermal pressure,
sourced by bulk gas motions and turbulence, reduces the thermal
pressure required to support the ICM against gravitational collapse
and thus the amplitude of the tSZ signal.  Similarly,
\citet{battaglia10} demonstrated that the inclusion of radiative
cooling, star formation and AGN feedback in hydrodynamical simulations
substantially lowers the tSZ power compared to simulations that omit
these processes.  Current predictions for tSZ power from models and
simulations are consistent with the upper limits derived from
observations.

These recent modeling
developments are supported by data from \Planck; when the models are
used to extrapolate from X-ray measurements to a predicted tSZ signal,
the predictions agree with \Planck SZ observations.  Agreement is
seen both in observations of single galaxy clusters
\citep{planck2011h, planck2011i} and via a stacking analaysis over a
broad range in X-ray luminosity down to masses as small as $M_{500}
\sim 5 \times 10^{13} M_\sun$ \citep{planck2011j}.

Current data provide no direct lower limits to the amplitude of tSZ
power due to a degeneracy with the kinetic SZ power spectrum
\citep{lueker10}.  The kinetic SZ effect arises due to the Doppler
Thomson scattering of CMB photons off of regions of ionized gas with
bulk peculiar velocities.  Upper limits on kSZ power set by
\citet{lueker10} and now substantially tightened by
\citet{shirokoff10}, are ruling out some models of patchy reionization.
It is useful to decompose the kSZ power into contributions arising
from an inhomogeneous transition from a neutral to ionized
inter-galactic medium, so called ``patchy reionization,'' and those
from the post-reionization era, the ``Ostriker-Vishniac'' (OV) effect.
The former is much more uncertain than the latter, and our best
knowledge of its amplitude comes directly from the upper limits in
\citet{shirokoff10}.  The OV power level has a current theoretical
uncertainty that we estimate to be about a factor of 2.  Despite its
low levels, kSZ power is a worrisome source of potential bias of
cosmological parameters since its spectral dependence is the same as
the primary CMB temperature anisotropies.

We expect that the only potentially significant extragalactic
contributions to polarization anisotropy are Poisson power from radio
sources and DSFGs.  A polarization analog for DSFG clustering could
only arise due to (unexpected) correlations between galaxies in the
polarization orientations of their emission.  Polarization signals arise
from scattering off of electrons in clusters and groups
\citep{sazonov99,carlstrom02,amblard05} and in reionized patches
\citep{knox98, santos03}, but these are also expected to be negligibly small.

In addition to developing and exploring the implications of an
extragalactic foreground model that takes into account recent
developments, we introduce a new approach to analyzing the
multi-frequency data.  We show how the complexities of our modeling
can be reduced to a fairly simple description of the contamination of
the estimates of CMB power spectra.  The contamination can be
described by just a few principal components whose amplitudes are
constrained by CMB--free linear combinations of the auto and
cross-frequency power spectra.

The outline of our paper is as follows. In Sec.~\ref{sec:modeling}
we describe our foreground models before describing our
fiducial models and surveys in Sec.~\ref{sec:fid} and \ref{sec:surv}
respectively. In Sec.~\ref{sec:method} we describe our general methodology
before detailing our principal component approach in
Sec.~\ref{sec:lincombo}.  We finally present our results in
Sec.~\ref{sec:res} and discuss them in Sec.~\ref{sec:disc}.

\newpage
\section{Modeling}
\label{sec:modeling}
\subsection{Emission from External Galaxies}
\label{sec:extgal}

External galaxies in the bands of interest are well approximated by power-law 
intensities $I_\nu \propto \nu^\alpha$, and divide fairly cleanly into 
those with spectral indices $\alpha < 1$ (radio galaxies) and those with
$\alpha > 1$ (DSFGs) \citep{vieira10}.

External galaxies lead to anisotropy via their discreteness, usually
modeled with a Poisson distribution, and also via correlations due to their 
tracing of the large-scale structure.  The Poisson fluctuations are
important for both radio galaxies and DSFGs, while clustering is only
significant for the dusty galaxies \citep{hall10}. 

The Poisson contribution depends on the brightness function, $dN/dS$,
via 
\be
C_\ell = \int_0^{S_c} dS\, S^2 \frac{dN}{dS}
\ee
where $S_c$ is the flux cut; map pixels with sources with $S > S_c$
are masked.  Clustering power, in contrast, scales approximately with the square of
the mean intensity, $I_\nu^2$, with 
\be
I_\nu = \int_0^{S_c} dS\,S \frac{dN}{dS}.
\ee
Although radio sources do cluster, their mean intensity at the relevant
frequencies is much smaller than for the DSFGs; sufficiently smaller
that their clustering power is negligible.

\subsubsection{Radio Galaxies}

From \citet{vieira10} we know the radio galaxies at 150\,GHz and
220\,GHz, at flux densities below 100 mJy, are described quite well by
the \citet{dezotti05} model\footnote{We also know from recent
  \Planck results \citep{planck2011m} that at brighter flux
  densities the deZotti model significantly over predicts the number
  counts.}.  This model has a brightness function that is approximately
a power-law $S dN/dS \propto S^{\gamma_R}$. This translates into
Poisson power which depends on the flux cut via $C_\ell \propto
S_c^{\gamma_R+2}$.  Due to the inhomogeneity of the \Planck sky
coverage, $S_c$ will vary significantly across the sky.  So
that these angular variations can be taken into account, we chose to
model the radio galaxies in terms of $dN/dS$ rather than $C_\ell$.

For frequency dependence, we assume a power law spectrum for each source
with spectral index $\alpha = \bar \alpha + \delta \alpha$ and the
departures from the mean $\langle \delta\alpha^2 \rangle = \sigma^2$ uncorrelated from source to source.  
With these assumptions our power spectra from radio sources are given by\footnote{In deriving this form we have used the identity that for a zero-mean Gaussian random variable $x$, $\langle \exp(-x^2) \rangle = \exp(-\langle x^2\rangle/2)$. This identity and its applicability in this context, was pointed out to us by Challinor, Gratton and Migliaccio.}
\begin{equation}
\label{eq:cl_rp}
C_\ell^R = C^{R,0} \left(\frac{S_c}{S_0}\right)^{\displaystyle \gamma_R + 2} \left(\frac{\nu \nu'}{\nu_0^2}\right)^{\displaystyle \left[ \alpha_R + \ln(\nu \nu'/\nu_0^2)\sigma_R^2/2 \right]}
\end{equation}

\subsubsection{Dusty Star-Forming Galaxies}

For the DSFGs, even the contribution to the shot noise falls 
well below the flux threshold \citep{hall10}, thus the
Poisson power is nearly independent of flux cut
and we choose to build our model in $C_\ell$ rather than $dN/dS$.
In that case, the DSFG Poisson contribution is given simply by
\begin{equation}
\label{eq:cl_dp}
C_\ell^D = C^{D,0} \left(\frac{\nu \nu'}{\nu_0^2}\right)^{ \displaystyle \left[\alpha_{D} + \ln(\nu \nu'/\nu_0^2)\sigma_{D}^2/2 \right]}
\end{equation}

A number of authors have considered the clustering of the infrared
background, starting with \citet{bond86,bond91}.  Further theoretical
investigation \citep{scott99, haiman00} was stimulated by the
detection of the infrared background in COBE data
\citep{puget96,fixsen98}, and the detection of bright
``sub-millimeter'' galaxies in SCUBA data \citep{hughes98}.  Subsequently,
the clustering has been detected at 160 microns \citep{lagache07}, at
250, 350 and 500 microns by the Balloon-borne Large Aperture
Submillimeter Telescope \citep[BLAST]{viero09} and at 217\,GHz
\citep{hall10, dunkley10}.  Recent \Planck measurements of the CIB 
\citep{planckcib} have extended to much larger angular scales than before at
217\,GHz, 353\,GHz, 545\,GHz, and 857\,GHz and recent {\it Herschel}
measurements \citep{amblard11} have tightened up the BLAST measurements and extended
them to smaller angular scales.  The field is {\em rapidly} evolving.

For the clustering, we assume the same model as in \citet{hall10},
extended to phenomenologically include the consequences of non-linear
clustering by including a multiplicative factor which is a power-law
in $\ell$ for $\ell>1500$. Additionally, the \citet{hall10} model has
a spectral index that varies with $\ell$, because the radial window
function varies with $\ell$.  This effect leads to corrections on the
order of only 1\% across the relevant frequency range so we ignore
it. Thus, the DSFG clustering power spectra are
given by

\be
\label{eq:cl_dc}
C_\ell^C = C^{C,0} \Phi_\ell^{\rm H10} 
\left(\frac{\nu \nu'}{\nu_0^2}\right)^{\displaystyle\alpha_{C}}
\left\{
\begin{array}{l l}
1 & \ell<1500 \\
{\displaystyle \left(\frac{\ell}{1500}\right)^{\displaystyle n_C}} & \ell>1500
\end{array}
\right.
\ee
where $\Phi_\ell^{\rm H10}$ is the \citet{hall10} clustering template.

Though the same sources generate both the Poisson power and the
clustering power, they are weighted differently, thus for our baseline
model we conservatively assume no relationship between the clustering
spectral index and the Poisson spectral index.

\begin{figure}
\centering
\includegraphics[width=3.5in]{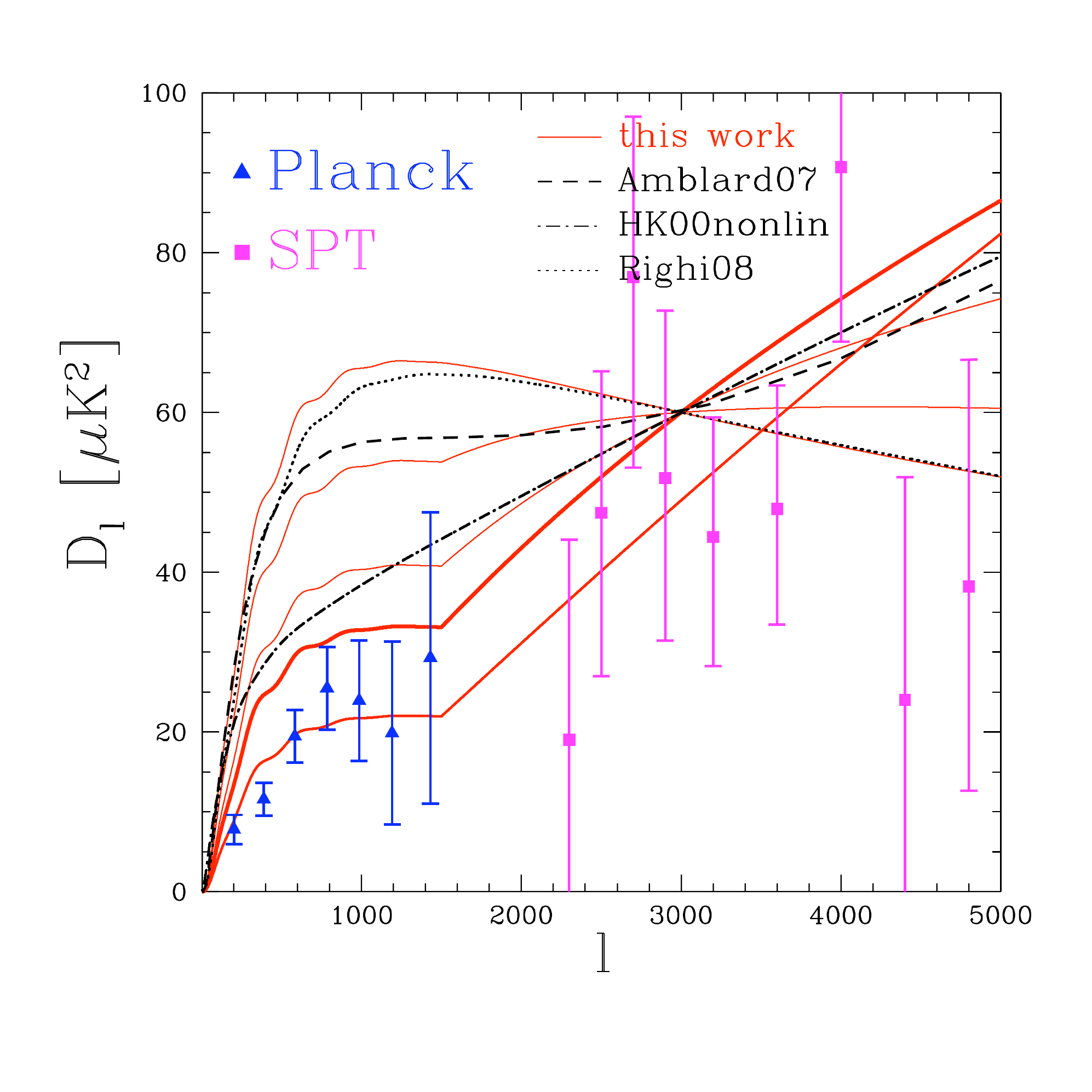}
\caption{Three model DSFG clustering auto-spectra at 217\,GHz (black),
 and approximations to them with our parameterized model (solid, red), 
all normalized (with one exception) at $\ell = 3000$.  Our fiducial model is the thickest
curve. Also plotted are estimates of the clusterng power from Planck \citep{planckcib} and SPT
\citep{hall10}.  For both sets of data points we have subtracted estimates of the Poisson
power from the reported total CIB power.  The lowest amplitude solid
(red) curve is the result of a ``by-hand'' adjustment of our model
parameters to fit the \Planck and SPT data. }
\label{fig:cibclustering}
\end{figure}

To gain some idea of the range of possible shapes of the DSFG
clustering power spectrum, we show a sampling of power spectra from
models in the literature in \fig~\ref{fig:cibclustering}.  They are all
normalized at $\ell = 3000$ to highlight similarities/differences in
shape. The models are the fiducial model from \citet{righi08}, the
$\beta = 0.6$ model from \citet{amblard07} and a non-linear version
of the model by \citet{haiman00}, hereafter HK00. Though they have similar shapes in the linear 
regime at large scales, then turn to a power-law behavior 
at high $\ell$, they result from very
different modeling assumptions.  \citet{righi08} associate the sources of infrared
light with starbursts triggered by mergers.  \citet{amblard07}
incorporate nonlinearities using a halo model.  For the `HK00nonlin'
curve, we used the luminosity densities for the fiducial model of HK00, assumed light is
a biased tracer of mass, and calculated the non-linear mass power spectrum using
the prescription by \citet{peacock96}. We will show that our phenomenological 
model is sufficient to reproduce these shapes with enough accuracy for \Planck cosmological parameter estimation. 

One result of the \Planck measurements, available only after our
calculations for this paper were completed, is that the CIB power
spectrum uncertainty at $\ell < 2000$ is now much smaller than before.
At least two of the three models shown in
\fig~\ref{fig:cibclustering} that guided our understanding of the
range of possible amplitudes have shapes that are inconsistent with
the combined \Planck and SPT data.  That range of possible amplitudes
is now given by the \Planck CIB power spectrum measurement
uncertainty.

\subsubsection{Polarization}

We expect polarized emission from the sources we consider to be
very small and uncorrelated from source to source.  For a collection
of sources with polarization fraction $f$, contributing a Poisson
temperature power spectrum of $C_\ell^{TT,P}$, we have
\bea
C_\ell^{EE} =C_\ell^{BB} & = & f^2 C_\ell^{TT,P} \nonumber\\
C_\ell^{TE} & = & f C_\ell^{TT,P}
\eea
We parameterize both radio source and DSFG contributions with the
above forms, with $f=f_D$ for DSFGs and $f=f_R$ for radio sources.

\newpage
\subsection{Thermal SZ Effect}
\label{sec:tsz}

The thermal SZ effect is a distortion of the CMB caused by inverse
Compton scattering of CMB photons off electrons in the high
temperature plasma within galaxy clusters. To first order, the temperature change
of the CMB at frequency $\nu$ is given by $\Delta T/T_{\rm CMB} (x_{\nu})
= f(x_\nu) y$, where $f(x_\nu) = x_\nu(\coth(x_\nu/2) - 4)$, $x_\nu = h\nu /
k_B T_{\rm CMB}$, and $y$ is the dimensionless Compton-y parameter
\begin{equation}
y = \left(\frac{k_B \sigma_T}{m_e c^2}\right)\int n_e(l)T_e(l) dl \;,
\end{equation}
where the integral is along the line of sight. $T_{\rm CMB}$ is the
CMB temperature, $n_e$ and $T_e$ are the number density and electron
temperature of the ICM, respectively.

The thermal SZ power spectrum can be calculated by simply summing up
the squared, Fourier-space SZ profiles, $\tilde{y}$, of all clusters:
\begin{equation}
C_{\ell}^{tSZ} =  f(x_\nu)^2\int dz {dV \over dz } \int d \ln M {dn(M,z) \over d \ln M}
\tilde{y}^2(M,z,\ell)
\label{eq:tsz_powerspec}
\end{equation}
where V(z) is the comoving volume per steradian and $n(M,z)$ is the
number density of objects of mass $M$ at redshift $z$. For the latter
we use the fitting function of \citet{tinker08}. $y(M,z,r)$ is the
projected radial SZ profile for a cluster of mass $M$ and redshift
$z$. Note that this calculation assumes that halos are not spatially
correlated; \citet{komatsu99} demonstrated that for $\ell > 1000$ the
two-halo (or clustered) contribution to the tSZ power spectrum is
several orders of magnitude smaller than the Poisson contribution
given by Eq. \ref{eq:tsz_powerspec}.

To calculate the thermal SZ signal we adopt the analytic intra-cluster
gas model presented in \citet{shaw10}. This model provides a
prescription for calculating the compton-y (or equivalently, thermal
pressure) profiles of hot gas in groups and clusters.  The model
assumes that gas resides in hydrostatic equilibrium in the potential
well of dark matter halos with a polytropic equation of state. The
dark matter potential is modeled by a Navarro-Frenk-White profile
\citep{navarro97} using the halo mass - concentration relation of
\citet{duffy08}. The model includes parameters to account for gas
heating via energy feedback (from AGN or supernovae) plus dynamical
heating via mergers. The stellar component of the baryon fraction in
groups/clusters is determined using the stellar mass fraction - total
mass relation observed by \citep{giodini09}. A radially-dependent
non-thermal pressure component of the gas is incorporated by
calibrating off the non-thermal pressure profiles measured in
hydrodynamical simulations \citep{lau10}. In total the model has four
free parameters relating to astrophysical processes in groups and
clusters. \citet{shaw10} explored the range in which these parameters
reproduce radial profiles and scaling relations derived from X-ray
observations of nearby groups and clusters.

To allow the astrophysical uncertainty to be marginalized over quickly
in our MCMC chains, we perform a principal component analysis (PCA)
described in Appendix \ref{app:pca}. A suite of 10,000 simulated power
spectra were created, each time randomly sampling from the input
astrophysical parameter distribution (with the cosmological parameters
fixed to their fiducial values described in Sec. \ref{sec:fid}). We
find that two principal components are sufficient to achieve 1\%
accuracy on the model power spectra.

\begin{figure}
\begin{center}
\includegraphics[width=3.5in]{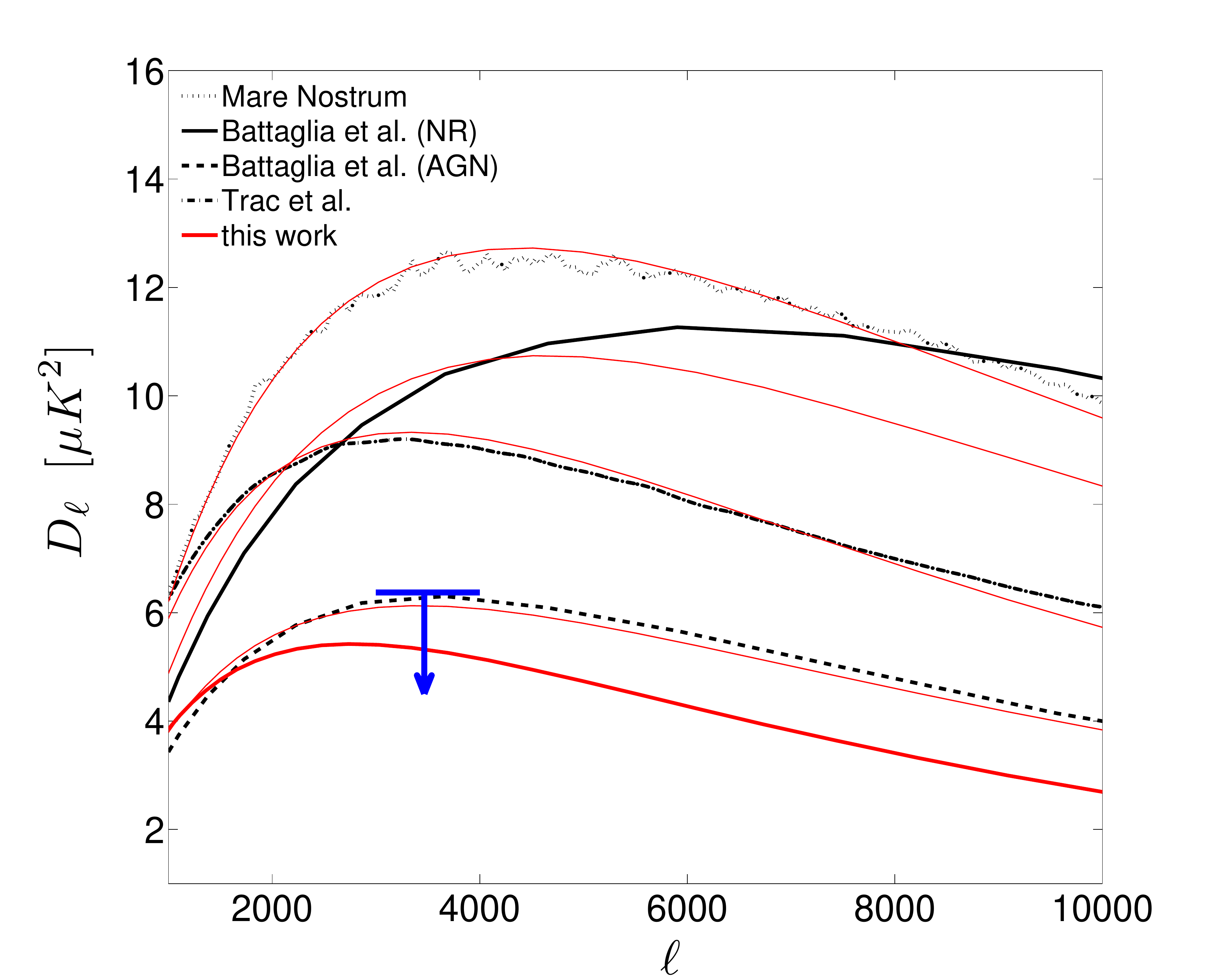}
\caption{Comparison between recent models and simulations of the tSZ effect (black lines)
  and fits of our PCA model to each (thin red lines).
  The thickest red line shows the fiducial tSZ power
  spectrum used in this work. All results are plotted at 146 GHz and
  are scaled to $\sigma_8 = 0.8$. The blue
  arrow shows the SPT 95\% confidence upper limit on thermal SZ
  power at $\ell = 3000$ \citep{shirokoff10}.}
\label{fig:tsz_pca_models}
\end{center}
\end{figure}

In \fig \ref{fig:tsz_pca_models} we plot the thermal SZ power spectrum
predicted by a number of recent simulations (black lines) as well as a
fit to each with our PCA model (red lines). The dotted line represents
the thermal SZ power spectrum measured from the Mare-Nostrum
simulation -- a non-radiative simulation run using the
smoothed-particle hydrodynamics code, Gadget-2. The black solid line
shows the results of the non-radiative simulation of
\citet{battaglia10} and the black dashed line the results of a rerun
of this simulation including radiative cooling, star-formation and
energy feedback. The dot-dashed line shows the `standard' tSZ model
from the simulations of\citet{trac10}. The thickest red lines represents
our fiducial thermal SZ model in this work. The blue point with
errorbars show the recent SPT constraint on the amplitude of thermal
SZ power at $\ell = 3000$. All models are plotted at 146 Ghz and have
been scaled to our fiducial cosmology.

Our PCA model can accurately reproduce all the simulations in Fig
\ref{fig:tsz_pca_models} other than the non-radiative simulation of
\citet{battaglia10} which peaks at much smaller angular scales than
the other simulations. We note that the \citet{shaw10} model inherently assumes that
some fraction of cluster gas has been converted to stars, whereas this
simulation did not include these processes. Turning off star formation
produces a power spectrum that peaks at smaller
scales.  In Section \ref{sec:model_sufficiency} we investigate the
bias on measured cosmological parameters when the \citet{battaglia10}
non-radiative template is used for the tSZ signal. We find that the
PCA will adapt sufficiently to prevent a bias in the measured
cosmological parameters.

The final step is to determine the cosmological scaling of the power
spectrum of our fiducial model so that the amplitude can be scaled
accordingly in our analysis. We find that the tSZ power spectrum is
principally sensitive to $\Omega_m$, $\Omega_b$, $n_s$, and
$\sigma_8$, with a particularly strong dependence on the latter. To
determine the scaling we simply evaluate our model varying each
cosmological parameter in the range $\pm 25\%$ of its fiducial value
while holding the other three fixed (at their fiducial value). We then
fit to the resulting power spectra, with our results summarized in
\tab \ref{tab:szscalings}.

\subsection{Kinetic SZ Effect}
The kinetic SZ effect is a temperature anisotropy that
arises from the Compton scattering of CMB photons off of electrons that have been
given a line-of-sight peculiar velocity by density inhomogeneities in the matter
field. We break up the kSZ into contributions from the post-reionization 
period and from a period of inhomogeneous ``patchy" reionization.

\subsubsection{Ostriker-Vishniac Effect}

\begin{figure}
\centering
\includegraphics[width=3.5in]{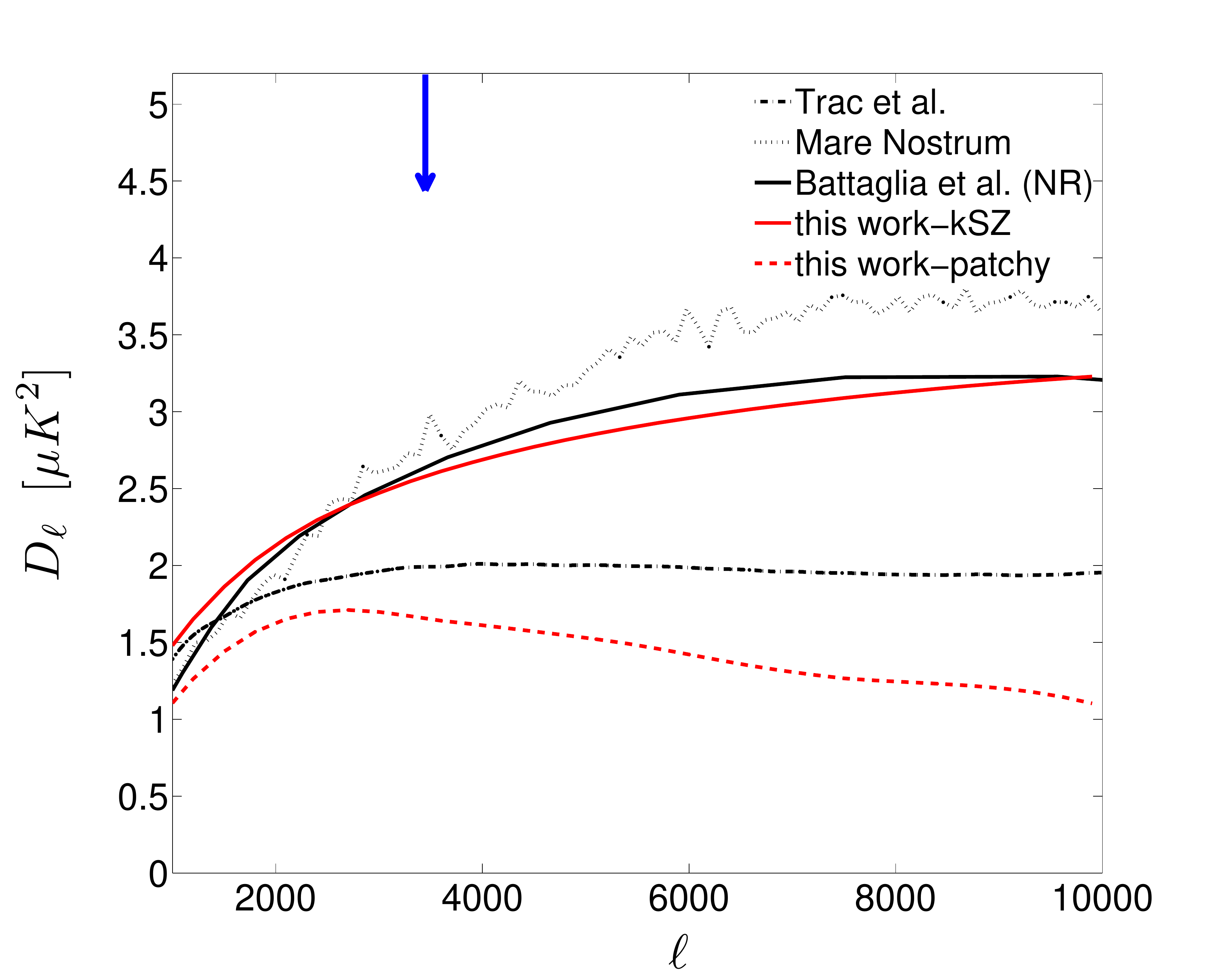}
\caption{Comparison between recent simulations of the kSZ effect
  (black lines) and our model. We consider contributions from the
  post-reionization kSZ effect (solid red line) and from patchy
  reionization (dashed red line). Note that the simulations plotted
  here assume homogeneous reionzation and thus do not include a patchy
  contribution. The SPT 95\% confidence upper limit for the kSZ power
  at $\ell = 3000$ is 6.5 $\mu {\rm K}^2$.}
\label{fig:ksz}
\end{figure}

When the density fluctuations which source electron velocities are in the linear 
regime the effect is known as the Ostriker-Vishniac (OV) effect, as derived in
\citet{ov86} and \citet{vishniac87}. The post-reionization kSZ effect can then be
modeled as the nonlinear extension of the OV effect as described
below.

We follow the analytic prescription given in \citet{hu00} which describes the angular 
power spectrum of the linear Vishniac effect as
\begin{equation}
C_{\ell}=\frac{\pi^2}{2\ell^5}\int d \chi D_A^3\left(
g\frac{\dot{G}}{G}\right)^2\Delta_{\delta_b}^{4}I_V.
\label{eq:lin_ov}
\end{equation}
where $\chi$ is the conformal time, $G$ is the cosmological growth
function, $D_A$ is the comoving angular diameter distance, $g$ is the
visibility function, $\Delta_{\delta_b}^2$ is the linear theory baryon density power
spectrum and $I_V$ represents the mode coupling of the linear density and
velocity fields:
\begin{equation}
I_V=\int_0^{\infty}dy_1\int_{-1}^1d\mu\frac{(1-\mu^2)(1-2\mu y_1)}{y_1^3y_2^5}
\frac{\Delta_{\delta_b}^2(ky_1)}{\Delta_{\delta_b}^2(k)}
\frac{\Delta_{\delta_b}^2(ky_2)}{\Delta_{\delta_b}^2(k)},
\end{equation}
with
\begin{eqnarray}
\mu&=&\hat{\mathbf{k}}\cdot\hat{\mathbf{k}}_1\nonumber\\
y_1&=&k_1/k\nonumber\\
y_2&=&k_2/k=\sqrt{1-2\mu y_1+y_1^2}.
\end{eqnarray}

Due to an incomplete treatment of
the effects of pressure feedback from baryons we slightly over-predict
the power on very small scales. As described in \citet{hu00}, in this
formulation we can consider the kSZ effect to be the nonlinear
extension of the linear Vishniac effect. This approximation requires
replacing the linear density power spectrum in \eq \ref{eq:lin_ov}
with its nonlinear extension while leaving the contribution from the
velocity power spectrum unchanged:
\begin{equation}
C_{\ell}^{kSZ,OV}=\frac{\pi^2}{2\ell^5}\int d \chi D_A^3\left(
g\frac{\dot{G}}{G}\right)^2\Delta_{\delta_b}^{2(NL)}\Delta_{\delta_b}^{2}I_V.
\end{equation}
For the nonlinear power spectra (NL) we utilize the HALOFIT
\citep{smith03} model. In this calculation of the kSZ effect we assume
that the nonlinear density fluctuations are uncorrelated with the bulk
velocity field in which they lie. \citet{zhang03} argue that this
approximation may not hold in highly nonlinear regimes where
contributions from the curl of the nonlinear velocity field may become
important however we neglect these corrections here.

As in the previous section, we find a power-law approximation for
estimating the kSZ power as a function of cosmological parameters. The
kSZ angular power spectrum was calculated under the full analytic
formulation for a large suite of WMAP7-allowed LCDM cosmologies. An
MCMC was then performed in the 6 dimensional fitting-function
parameter space and best-fit marginalized values were found and are
listed in \tab \ref{tab:szscalings}

In Figure \ref{fig:ksz} we compare our calculation of the kSZ
power spectrum (solid red line) with that measured from recent
simulations (black lines). As in Figure \ref{fig:tsz_pca_models}, we
plot the power spectrum predicted by the Mare-Nostrum simulation
(dotted), the `standard' model of \citet{trac10} (dot-dashed) and the
non-radiative hydrodynamical simulations of \citet{battaglia10}
(solid). The red dashed line shows our model for the contribution to
the kSZ signal from inhomogeneous reionization (not included in the
other lines), as described in the following section.

\begin{table}
\begin{center}
\caption{SZ Cosmological Scaling}
\begin{tabular}{ c c c c c c c }
\hline\hline
$\ell$ & $A_{\rm OV}$ [\muK2] & $n_s$ & $\Omega_b$ & 
$\Omega_c$ & $\sigma_8$ & $\tau$ \\
 & & 0.96 & 0.045 & 0.22 & 0.8 & 0.09 \\ [0.5ex]
\hline
500 & 1.18 &-1.44 & 1.83 &-1.06 & 4.36 & 0.25 \\ 
1000 & 1.81 &-1.36 & 1.91 &-1.13 & 4.82 & 0.24 \\ 
2000 & 2.64 &-0.94 & 1.96 &-1.12 & 5.26 & 0.22 \\ 
3000 & 3.06 &-0.45 & 1.94 &-1.03 & 5.38 & 0.20 \\ 
4000 & 3.33 &-0.22 & 1.96 &-1.04 & 5.54 & 0.17 \\ 
5000 & 3.53 &-0.03 & 1.98 &-1.05 & 5.66 & 0.15 \\ 
6000 & 3.67 & 0.13 & 2.00 &-1.06 & 5.75 & 0.13 \\ 
7000 & 3.78 & 0.27 & 2.01 &-1.07 & 5.83 & 0.12 \\ 
8000 & 3.87 & 0.38 & 2.02 &-1.08 & 5.89 & 0.10 \\ 
9000 & 3.94 & 0.49 & 2.03 &-1.09 & 5.94 & 0.09 \\ 
10000 & 4.00 & 0.58 & 2.04 &-1.10 & 5.99 & 0.08 \\
\\
\hline\hline
$\ell$ & $A_{\rm tSZ}$ [\muK2] & $n_s$ & $\Omega_b$ & 
$\Omega_m$ & $\sigma_8$ & $h$\\
 & & 0.96 & 0.045 & 0.265 & 0.8 & 0.71\\ [0.5ex]
\hline
500 & 2.06 &-1.01 & 2.41 & 0.69 & 8.57 & 1.30 \\ 
1000 & 3.59 &-0.75 & 2.45 & 0.63 & 8.49 & 1.43 \\ 
2000 & 4.86 &-0.36 & 2.52 & 0.53 & 8.40 & 1.65 \\ 
3000 & 5.04 &-0.08 & 2.57 & 0.48 & 8.36 & 1.73 \\ 
4000 & 4.82 & 0.15 & 2.62 & 0.44 & 8.34 & 1.88 \\ 
5000 & 4.44 & 0.31 & 2.66 & 0.42 & 8.33 & 2.03 \\ 
6000 & 4.03 & 0.49 & 2.70 & 0.39 & 8.32 & 2.13 \\ 
7000 & 3.62 & 0.58 & 2.73 & 0.38 & 8.32 & 2.18 \\ 
8000 & 3.24 & 0.77 & 2.78 & 0.36 & 8.32 & 2.27 \\ 
9000 & 2.89 & 0.87 & 2.80 & 0.35 & 8.33 & 2.32 \\ 
10000 & 2.59 & 0.96 & 2.83 & 0.34 & 8.33 & 2.37
\end{tabular}
\label{tab:szscalings}
\end{center}
\tablecomments{The $\ell$-dependent power-law cosmological scalings for the Ostriker-Vishniac effect and the thermal SZ effect. The numbers immediately below the cosmological parameters are the pivot points for the power law, and the numbers in the table are the power-law indices. For example, the top row says that for the OV effect, $D_{500} = 1.18$\muK2$\,(n_s/0.96)^{-1.44}(\Omega_b/0.045)^{1.83} ...$}
\end{table}

\subsubsection{Patchy Reionization}
We also consider the contribution to the kSZ power from
inhomogeneous reionization \citep{gruzinov98, knox98, hu00,
  zahn05, mcquinn05, iliev07}. Simulations and
analytic models of HII bubble formation both indicate
that the first galaxies and quasars were
highly clustered and led to gradual reionization in ``bubbles" that
quickly grew to sizes of several Mpc \citep{zahn10}.

In our estimates we use the analytic Monte-Carlo ``FFRT" model of
\citet{zahn10}. It has been shown to agree well with the most
sophisticated radiative transfer simulations on scales of 100 comoving
Mpc/h, while having the added advantage of allowing the modeling of
arbitrarily large volumes (the analytic scheme is about 4 orders of
magnitude faster, at a given dynamic range, than radiative
transfer). This is especially important for kSZ, since large scale
velocity streams lead to the bulk of the signal. Our particular
template (dashed red line in \fig \ref{fig:ksz}) was calculated in a 
1.5 Gpc/h cosmological volume where x- and
y-axes correspond to roughly 15 degrees on a side and z-axis
corresponds to redshift, with a median redshift of 8.
We shift this template left-right logarithmically by a ``patchy shift" parameter 
$R_P$; that is, the power spectrum for a given shift is related to the fiducial $R_P=1$ spectrum by,
\begin{equation}
C_\ell = C^{fid}_{R_P\times\ell}
\end{equation}
$R_P$ is to be thought of as scaling the size of the bubbles, and is, to good 
approximation, proportional to the duration of the patchy 
phase. The timing of reionization has a secondary small effect on the shape 
and amplitude, which we 
neglect here.

\subsection{tSZ-DSFG Correlation}

It is reasonable to expect some correlation between the DSFG clustering and tSZ components since they both trace the same underlying dark matter distribution, with significant overlap in redshift. Simulations which associate emission with individual cluster member galaxies predict anti-correlations---DSFGs fill in SZ decrements at frequencies below 217\,GHz considered here---on the order of tens of percent, with a correlation coefficient nearly independent of scale \citep{sehgal10}. This effect was explored in \citet{shirokoff10} which found correlation consistent with zero but with significant uncertainty due to degeneracies with the tSZ and kSZ components.

Assuming a fixed correlation $r_{tSZ,C}$, the total power spectrum isn't simply the sum of the tSZ and DSFG clustering terms, but must also include a term given by,
\begin{equation}
C_{\ell,\nu\nu'} = r_{tSZ,C} \left( \sqrt{C^C_{\ell,\nu\nu}C^{tSZ}_{\ell,\nu'\nu'}} +
\sqrt{C^{tSZ}_{\ell,\nu\nu}C^{C}_{\ell,\nu'\nu'}} \right)
\end{equation}
Note that this effect can be larger than either component individually in cross spectra between frequencies in which each component is large. For example, even with only moderate levels of correlation, the $217\times70$ GHz foreground contribution would be dominated by this correction because of the large DSFG power at 217\,GHz and tSZ power at 70\,GHz. 

\subsection{CMB}

For the primary CMB signal itself, we use an 8-parameter model which includes the baryon density $\Omega_b { h^2}$, the density of cold dark matter $\Omega_c { h^2}$, the optical depth to recombination $\tau$, the angular size of the sound horizon at last scattering $\Theta$, the amplitude of the primordial density fluctuations $\ln[10^{10} A_s]$, the scalar spectral index $n_s$, the dark energy equation of state parameter $w$, and the tensor-to-scalar ratio $r$. Freeing $w$ opens up the ``geometric degeneracy" which is typically broken by adding an external dataset such as supernovae data. Rather than do this, we simply put a $\pm 0.3$ Gaussian prior on $w$ to reasonably constrain the chain while allowing it to explore the parameter space.

Because we are interested in the simplest description of how the foregrounds affect cosmological parameters, we do not consider extensions to our model such as a running spectral index, non-flat universes, non-standard effective number of neutrino species, or a difference in primordial helium from standard big bang nucleosynthesis. Due to the small angular scales where they affect the CMB anisotropy, it is possible that such parameters could be even more degenerate with the foregrounds than the ``vanilla" set we consider. 

For quick and highly accurate CMB calculations during our MCMC chains, we use a PICO 
\citep{pico} interpolation of a training set generated by CAMB \citep{camb}. Though more 
recent recombination codes exist, we expect the one used in our training set to be 
sufficient for our forecasting. Additionally, the training set includes the 
option of a non-linear lensing contribution described in \citet{challinor05} which we use.

\section{Fiducial Model and Current Constraints}
\label{sec:fid}

For our forecasting, we create simulated power spectra (henceforth the ``simulated data") using the model described in the previous sections. We pick one single set of model parameter values, called the ``fiducial values" or the ``fiducial model" in general, which will is the baseline for the different cases of simulated data which we consider. The model used to analyze the simulated data, which generally contains small changes relative to the fiducial model, will be called the ``analysis model."

In \tab \ref{tab:params} we summarize all of the parameters 
in our model, the naming convention, and their fiducial
values.  The fiducial values are chosen to be consistent with current
cosmological constrains from WMAP7 \citep{komatsu10}, and with
constraints on the foreground components from ground-based data such
as SPT and ACT. In the following paragraphs, we describe the method used
to arrive at our fiducial model.

Because the expected SZ power depends strongly on
cosmology, special care was taken so that our fiducial SZ power and
cosmology agree. To achieve this, we use the constraint from
\citet{lueker10} on the linear combination $\mathcal{D}_{tSZ} + .46
\times \mathcal{D}_{kSZ} = 4.2 \pm 1.5$ \muK2 (at $\ell=3000$ and
153\,GHz) along with the cosmological scalings in \tab
\ref{tab:szscalings}. We then importance sample the WMAP7
$\Lambda$CDM+TENS\footnote{Available at {\rm
    http://lambda.gsfc.nasa.gov/}} chain by calculating at each step
the expected SZ (kinetic and thermal) power assuming no theory
uncertainty, then applying the prior from \citet{lueker10}.
The new best fit point in the post-processed chain mainly shifts SZ power up
relative to best fit SPT value and $\sigma_8$ down relative to the best
fit WMAP7 value. All other cosmological parameters are also affected
(at a smaller level), and their new mean values form our fiducial
cosmology, which remains 1-$\sigma$ consistent across all parameters
with WMAP7.

For the radio sources, the tightest constraints on the expected \Planck power come from the \citet{vieira10} catalog which contains sources in the decade of brightness just below the \Planck flux cut. Fitting a \citet{dezotti05} model to the data yields the values listed in \tab \ref{tab:params}, notably radio Poisson power of 133\,\muK2 at 143\,GHz assuming a 330 mJy flux cut.

Since the DSFG Poisson contribution is nearly independent of flux cut, 
we expect the same Poisson power in \Planck maps as in SPT maps,
adjusting only for bandpass differences. We get our fiducial value for \Planck 
Poisson power at 143\,GHz by extrapolating in frequency from the best-fit value of the SPT 150\,GHz power as given in \citet{shirokoff10}. Our fiducial values for $\alpha_D$ and $\alpha_C$ also come from the best-fit values 
in \citet{shirokoff10}.  We set $\sigma_D$ to 0.4 following the arguments
in \citet{knox04}, although it is not yet well constrained by observations.  We adopt a clustering tilt $n_C = 1$ so that it (roughly) has the shape
expected at small scales due to the the observed clustering properties
of high-redshift galaxies.  As argued by \citet{scott99}, the observed clustering
properties of $z \sim 3$ Lyman break galaxies, namely an angular
correlation function proportional to $\theta^{-0.9}$
\citep{giavalisco98}, correspond to $D_\ell \propto \ell^{1.1}$. 
Since we multiply the power-law by the linear theory
template, this is similar (at $\ell > 1500$) to the power-law only $D_\ell
\propto \ell^{0.8}$ shape used as baseline models in both \citet{dunkley10} and \citet{shirokoff10}. Although ruled out by the \Planck data, our
fiducial model is at least closer to the measurements than
all of the other models plotted in \fig \ref{fig:cibclustering}. 
 The agreement is sufficient for our
purposes here, though we will certainly be updating our CIB
modeling in the near future.

Following \citet{battye10} which found a mean fractional polarization
of 4.5\% at 86\,GHz (and varying weakly with frequency) for the WMAP
point source catalog \citep{wright09}, we adopt a fiducial value of $f_R = 0.05$.  For
DSFGs we expect an even smaller level of average polarization
fraction.  Polarized dust emission arises due to alignment of grains
in interstellar magnetic fields. We somewhat arbitrarily set $f_D =
0.01$ for our fiducial model which is consistent with the finding
that, in our own galaxy, the coherence length for magnetic fields is
much smaller than the extent of the dust emission \citep{prunet98}.

\begin{figure*}
\centering
\begin{tabular}{c}
\includegraphics[width=5in]{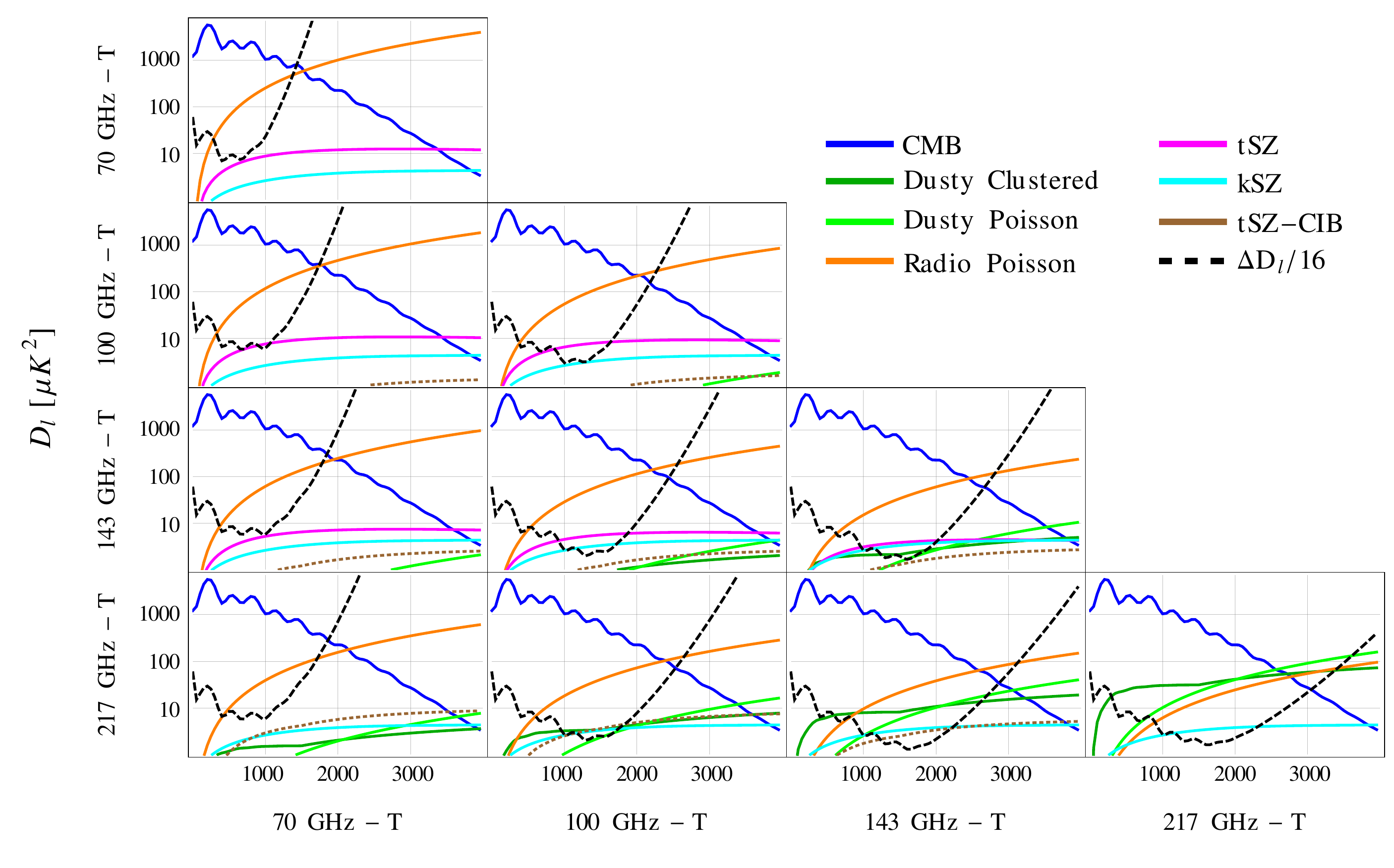}
\\
\includegraphics[width=5in]{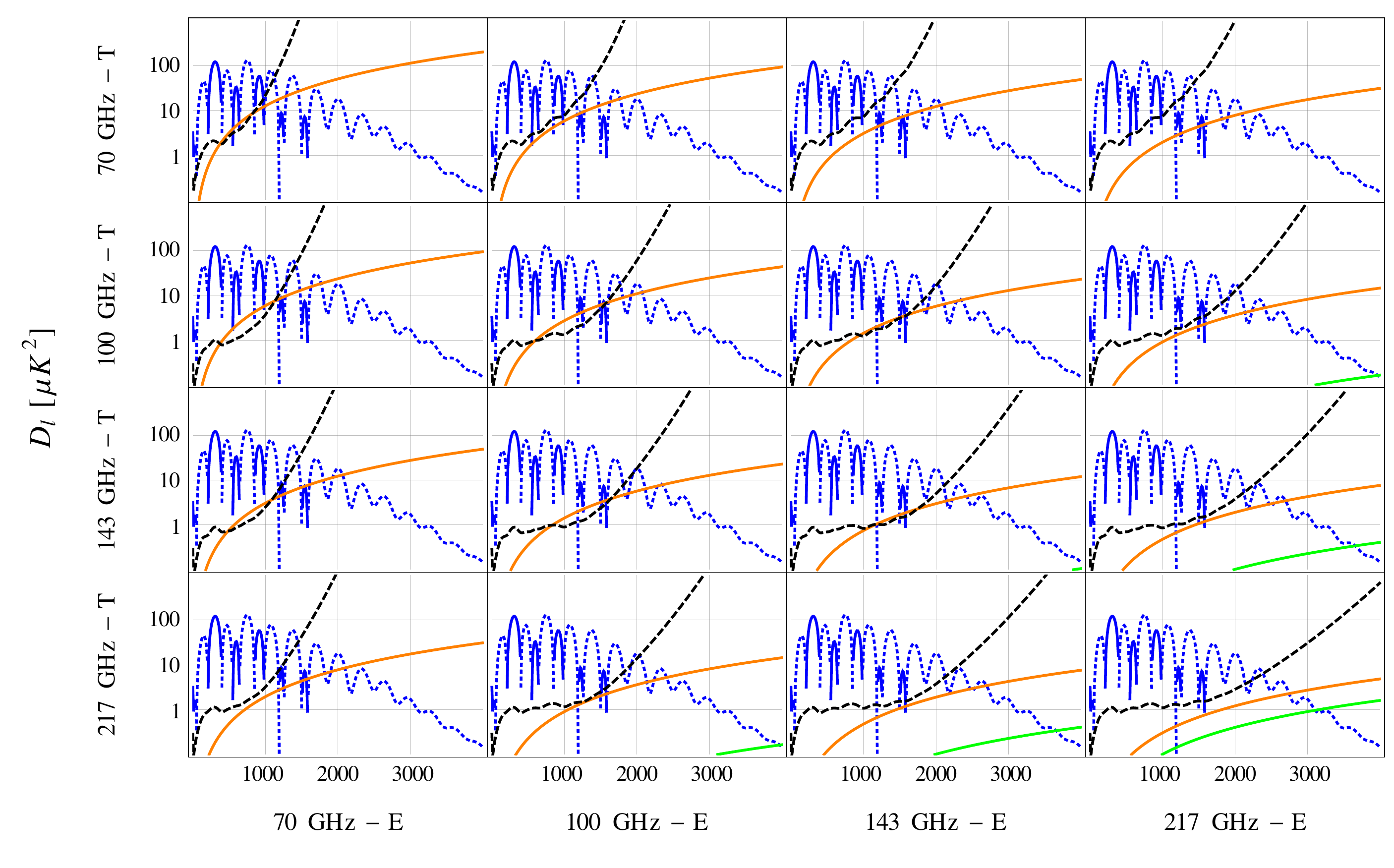}
\\
\includegraphics[width=5in]{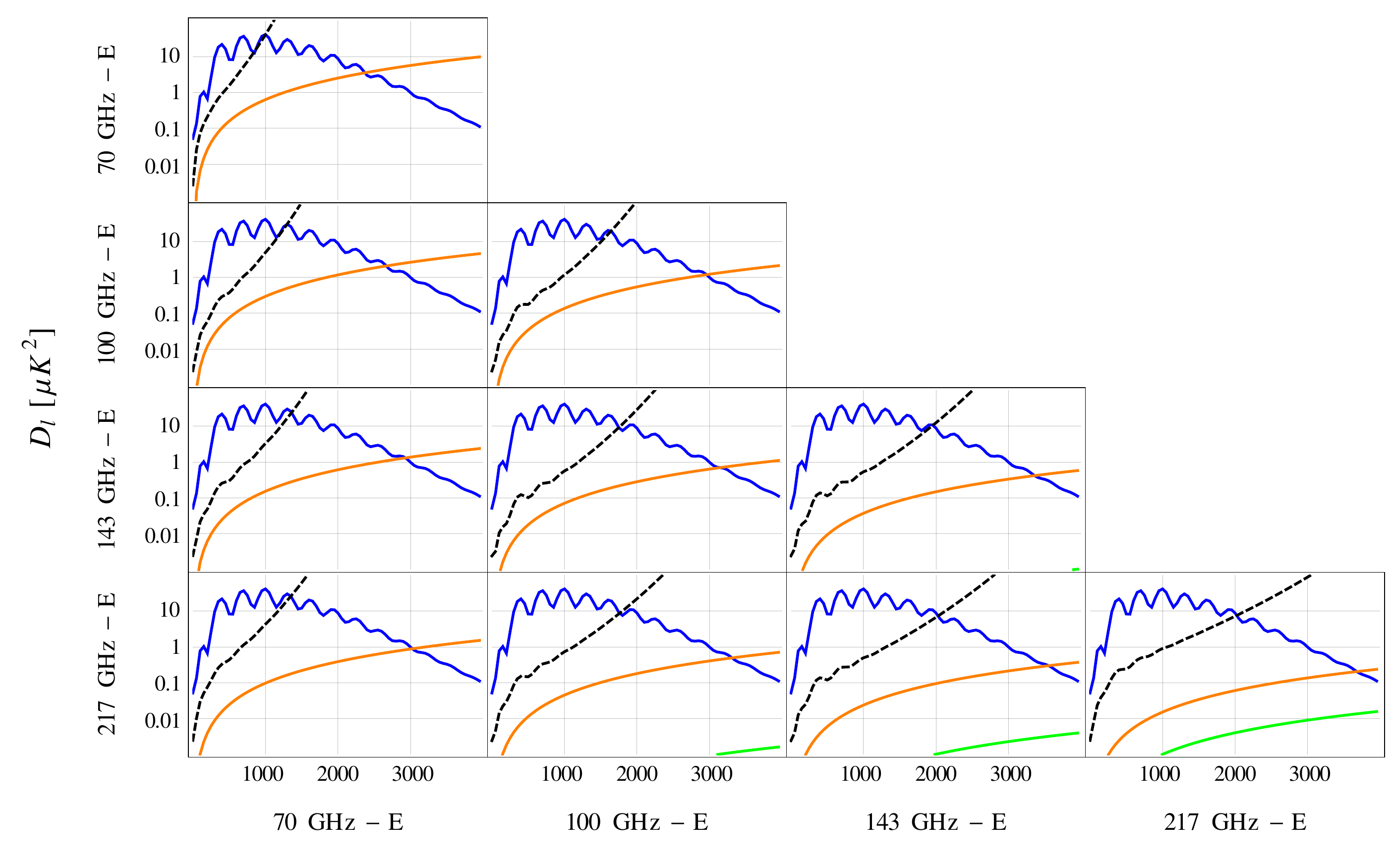}
\end{tabular}
\caption{All 36 power spectra which can be formed from \Planck 70\,GHz--217\,GHz temperature and E-mode polarization, and the prediction of our fiducial model for the CMB and foreground power in each of them (with the exception of the tSZ-DSFG correlation which is shown at 30\% instead of its fiducial value of 0\%). The black dashed line shows the errors bars for $\ell$-bins of width $\Delta\ell=256$. Dotted lines indicate negative power.}
\label{fig:grid}
\end{figure*}

\fig \ref{fig:grid} shows the fiducial CMB and foreground contribution to \Planck TT, TE, and EE power spectra (with the exception of tSZ-DSFG correlation which is plotted at 30\% rather than its fiducial value of 0\%).

\begin{table}
\centering
\caption{Survey Properties}
\begin{tabular}{c|c|c|c}
\hline\hline
Band & T (E/B) & Beam & Notes\\
(GHz) & ($\mu K$-arcmin) & (arcmin) \\ \hline
\textbf{\Planck} & & \\
70 & 177 (253) & 14 & $f_{sky}=70$\%\\
100 & 61 (98) & 10 & $S_{c}=330$\,mJy\\
143 & 42 (80) & 7.1 \\
217 & 64 (132) & 5 \\ \hline
\textbf{\Ground-Deep} & & \\
90 & 53 & 1.6 & $f_{sky}=100$\,deg$^2$\\
150 & 13 & 1.15 & $S_{c}=6.4$\,mJy\\
220 & 35 & 1.05 \\ \hline
\textbf{\Ground-Wide} & & \\
90 & 53 & 1.6 & $f_{sky}=1000$\,deg$^2$ \\
150 & 18 & 1.15 & $S_{c}=6.4$\,mJy \\
220 & 80 & 1.05 \\ \hline
\end{tabular}
\label{tab:surveyprop}
\tablecomments{Instrument properties used to generate simulated power spectra. The beam width is given as a full-width-half-max. $S_c$ refers to the flux cut above which brighter sources are masked out.}
\end{table}

\section{Survey Properties}
\label{sec:surv}

We consider simulated \Planck data in the four
bands between 70\,GHz and 217\,GHz. These are chosen because they
contain nearly all of the significant CMB information. Though the neglected channels place 
little extra constraints on the CMB, they are crucial for understanding and 
cleaning the foregrounds. We consider their effect implicitly by testing limits such as
lowered Poisson power amplitudes, or fixed DSFG clustering
shapes. Additionally, we also consider the benefit of higher resolution 
ground-based data, which we model after SPT 90\,GHz, 150\,GHz, and 220\,GHz 
channels. We divide the data into two fields: a 100\,deg$^2$ ``deep" field
and a 1000\,deg$^2$ ``wide" field. We henceforth refer to these two 
datasets as \Ground-deep and \Ground-wide. The depths, sky coverage,
and flux cuts used in our forecasting are summarized in \tab \ref{tab:surveyprop}.

Our simulated data take the form of auto and cross spectra from as
many bands as are present for a given patch of sky. The four \Planck frequency channels 
form 10 TT, EE, and BB and 16 TE power spectra, 
with an additional 18 TT power spectra from the
three extra frequencies in regions of \Ground overlap\footnote{In general 
$N$ frequency channels can be used to create $N(N+1)/2$ power
spectra of type TT, EE, and BB, and $N^2$ of type TE}. We do
not assume overlap between \Ground deep and wide, nor do we form cross
spectra between \Ground temperature and \Planck polarization as these
are expected to be a very small contribution to the CMB and foreground
information. \Planck BB polarization is also ignored except in one
test case where we find its impact is minimal on our cosmological parameterization.

We simulate power spectrum assuming a uniform masking threshold across the sky. The only exception is in the case of \Planck and \Ground overlap. For such patches of sky, we assume \Planck maps can be masked using a point source mask from the higher resolution \Ground data. Thus, for the overlap areas, even the \Planck auto spectra will have greatly reduced radio Poisson power. 

The non-zero width of frequency bandpasses creates a different effective frequency for each 
component in each band. For components with uncertain spectral shapes, the variation in
effective frequency leads to percent level corrections which can be neglected.
In this paper,
values quoted from ground-based experiments are normalized
at, and explicitly cite, the corresponding effective frequency. For \Planck,
it is sufficient for our forecasting purposes to ignore this and use
the nominal band centers for all components.
 
\begin{table*}
\caption{Summary of Model Parameters}
\begin{center}
\begin{tabular}{ c | l | l | l }
\hline\hline
Parameter & Fiducial Value & Current Constraints (1$\sigma$) & Definition
\\ [.5ex]
\hline
\textbf{Cosmological} & & \\
$\Omega_b h^2$ & .022565 & .00073 & Baryon density\\
$\Omega_c h^2$ & .10709 & .0063 & Cold dark matter density \\
$\Theta$ & .010376 & .000029 & Angular size of the sound horizon at last scattering  \\
$\tau$ & .0799 & .015 & Optical depth to reionization \\
$w$ & -1 & .13 & Dark energy equation of state parameter \\
$n_s$ & .9669 & .014 & Scalar spectral index \\
$\ln(10^{10} A_s)$ & 3.1462 & .045 & Scalar amplitude \\
$r$ & .13 & $<$.36 (95\%) & Tensor-to-scalar ratio \\ [.5ex]
\hline
\textbf{Dusty Poisson} & & \\
$\alpha_{D}$ & 3.8 & 0.35 & Spectral index \\
$\sigma_{D}$ & .4  & & Spectral index intrinsic spread\\
$\mathcal{D}_{D}$ & $5.9 \,$\muK2  & 0.8 & Amplitude at $\ell=3000$, $\nu=143$\,GHz\\
$f_{D}$ & .01 & & Dusty polarization fraction \\ [.5ex]
\hline
\textbf{Radio Poisson} & & \\
$\alpha_{R}$ & -.5  & 0.1 & Spectral index \\
$\sigma_{R}$ & .1  & $<0.6$ (95\%) & Spectral index intrinsic spread \\
$\mathcal{D}_{R}$ & $133 \,$\muK2 & 27\,\muK2 & Amplitude at $\ell=3000$, $\nu=143$\,GHz, $S_c=330$\,mJy\\
$\gamma_{R}$& -.8 & 0.1 & Brightness function power law index \\
$f_{R}$ & .05 & & Polarization fraction \\ [.5ex]
\hline
\textbf{Dusty Clustered} & & \\
$\alpha_{C}$ & 3.8 & 0.4 & Spectral index \\
$\mathcal{D}_{C}$ & $3.9 \,$\muK2 & 1.2\,\muK2 & Amplitude at $\ell=3000$, $\nu=143$\,GHz\\
$n_{C}$ & $1$ & & Nonlinear tilt \\
\hline
\textbf{SZ Effects} & & \\
$\mathcal{D}_{tSZ}$ & $4.3 \,$\muK2 & $<6.8$\,\muK2 (95\%) & tSZ amplitude at $\ell=3000$, $\nu=143$\,GHz \\
$\mathcal{D}_{kSZ,OV}$ & $2.7 \,$\muK2 & $<6.5$\,\muK2 (95\%)& OV amplitude at $\ell=3000$ \\
$\mathcal{D}_{kSZ,P}$ & $1.5 \,$\muK2 & $<6.5$\,\muK2 (95\%)& Patchy amplitude at $\ell=3000$ \\
$R_{P}$ & 1 & & Patchy shift \\
\hline
\textbf{Correlations} & & \\
$r_{tSZ,C}$ & 0 & & Correlation between tSZ and DSFGs at $\ell=3000$
\end{tabular}
 \normalsize
 \tablecomments{A summary of the parameters in our model. The fiducial values generate our simulated data. The current constraints column gives the 1$\sigma$ constraints on our model given WMAP power spectra and radio source counts, SPT power spectra and radio/DSFG source counts, and ACT power spectra. Note that due to the process by which the fiducial values were chosen (Sec.~\ref{sec:fid}) they are not necessarily the most likely values given current data; they are, however, totally consistent with the most likely value to within 1$\sigma$.}
\end{center}
\label{tab:params}
\end{table*}

\section{Forecasting Methodology}
\label{sec:method}

The analysis of the simulated data assumes perfectly known Gaussian beams, no 
calibration uncertainty, isotropic noise, and ignores the effects of mode-mode coupling on 
the cut sky. While these assumptions are not sufficient for modeling real data, we expect 
them to be adequate for our purpose of modeling the extragalactic foregrounds, and 
understanding their importance on cosmological parameter biases and statistical errors. 

Under these assumptions, the so called ``pseudo power spectrum" which includes both signal and noise is given by,
\begin{equation}
C_{ij,\ell} = C^S_{ij,\ell} + \delta_{ij} w^{-1} \exp(\ell^2 \sigma_b^2/2)
\end{equation}
Here $i$ and $j$ each label one of the maps, and the noise is
parameterized by the weight per solid angle $w$, and the beam width in
radians $\sigma_b$.  
To calculate the error on our estimates of the signal power spectrum,
we use a Fisher matrix approximation which is 
nearly exact in the $\ell$ range we consider.  The resulting ``bandpower covariance" is
\begin{eqnarray}
\label{eq:cov}
\Sigma_{(ij)(kl)} &\equiv& \left< ( \widehat C^S_{ij} - C^S_{ij} )( \widehat C^S_{kl} - C^S_{kl} ) \right> \nonumber
\\
&=&\frac{1}{(2l+1)f_{sky}} \left( C_{il} C_{jk} + C_{ik} C_{jl} \right)
\end{eqnarray}
where we have suppressed the $\ell$ dependence for notational
simplicity.  We use the covariance to form the likelihood as a function of parameters $\theta$,
\begin{equation}
-2 \ln \mathcal{L}(\theta) = \left[ C^S_{ij}(\theta) - \widehat C^S_{ij} \right] \Sigma^{-1}_{(ij)(kl)} \left[ C^S_{kl}(\theta) - \widehat C^S_{kl} \right]
\end{equation}
Note we have neglected the normalization term since it does not vary with $\theta$. 

Our simulated data are the mean expected power spectra; i.e., they do
not include a sample of the errors from the bandpower covariance
matrix.  Leaving out these fluctuations has the benefit of making the best fit $\chi^2$ equal to
exactly 0 (as long as our analysis model and simulation model are the
same) and has no affect on our forecasting abilities.

Tests we performed showed that a Gaussian propagation of uncertainty
from the $C_\ell$'s to the model parameters can be insufficiently
accurate due to the highly non-Gaussian shape of the foreground
parameter posterior likelihoods. Instead, we run a full Markov-Chain
Monte-Carlo (MCMC) analysis using a custom
multi-frequency extension to CosmoMC \citep{cosmomc}. The code for
this extension along with chains for the results quoted in this paper
are available online\footnote{http://student.physics.ucdavis.edu/$\sim$millea/data/millea2011}.

\section{Compression to a CMB Power Spectrum Estimate}
\label{sec:lincombo}
Before getting to our results, it is useful to explore the foreground contamination in a more model-independent manner, motivated by two drawbacks of our procedure. First, there is a large amount of data one must work with---our bandpower covariance matrix at each $\ell$ is 46$\times$46 for \Planck and 64$\times$64 for \Planck+\Ground. Second, 17 foreground parameters must be marginalized over, and if one wanted to examine constraints on a new cosmological model, the whole procedure would have to be repeated.  Here we present a procedure for compressing all the power spectra to 1) a single CMB estimate and 2) a low dimensional parametrization of the residual foregrounds in this estimate.  We describe the procedure here for temperature-only power spectra with errors that are uncorrelated from multipole to multipole.  The generalization to include $\ell$ to $\ell$ correlations and polarization is in Appendix \ref{app:lincombo}.

Given $N$ power spectra (for example the 10 TT spectra we consider for \Planck), 
we would like to split our data up into $N-1$ linear combinations of
power spectra that have no sensitivity to the CMB and then find the
remaining linear combination that contains CMB and whose errors are
uncorrelated with those of the $N-1$.  With
this split made, we can then derive our foreground constraints using the
CMB--free linear combinations.  Doing so means foreground
constraints can be made independent of our modeling of the CMB (other than the
assumed frequency dependence).  

We use the $N-1$ CMB--free linear combinations to find the constraints
they place on our foreground model parameters via MCMC.  For each
point in the chain we can determine the contribution to the CMB linear
combination.  We sample over all these contributions to find the mean
contribution and fluctuations about that mean.  We find a
low-dimensional description of the fluctuations via a principal
component decomposition.  

\begin{figure}
\centering
\includegraphics[width=3in]{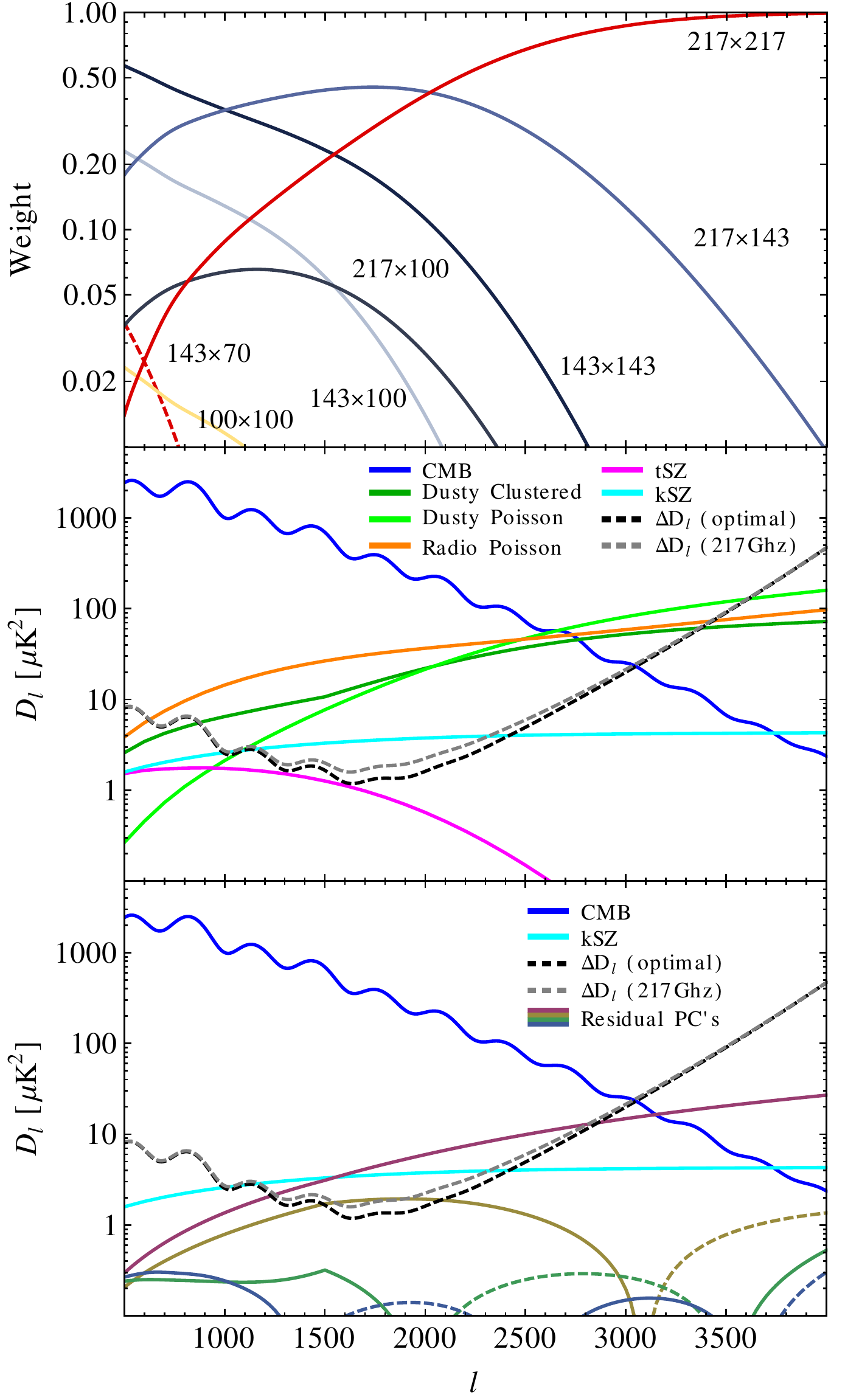}
\caption{\textit{Top}: The $\ell$-dependent weightings which form the CMB linear combination (Eq. \ref{eq:wcmb}). All possible auto/cross spectra from \Planck channels in \tab \ref{tab:surveyprop} were considered. Dashed lines indicate negative weight. \textit{Middle}: The mean foreground contribution to the CMB linear combination for our fiducial model. Note for example that tSZ (purple) is not present at high $\ell$ because only 217\,GHz is used there. The dominant non-Poisson component for the $\ell$-range where \Planck is most sensitive is the DFSG clustering. \,\textit{Bottom}: Principal components of foreground residuals (constrained by the CMB--free linear combinations) with amplitudes
set to 1-$\sigma$. Note that we only need two principal component amplitudes to be accurate to $>1$\muK2. (The errors in bin widths of $\Delta \ell=256$ for both the CMB linear combination and for 217\,GHz alone are plotted as dashed lines in the bottom two plots.)}
\label{fig:lincombo}
\end{figure}

\subsection{Splitting the power spectra into CMB--free and a CMB
  estimate}

Let us begin by first considering arbitrary linear combination of
the power spectra, 
\begin{equation}
\tilde C^\mu = \sum_i w^\mu_ i C_i
\end{equation}
where $C_i$ are the $i = 1$ to $N$ power spectra.  The weightings we consider will be
$\ell$-dependent; the lack of any labeling by $\ell$ is solely for
notational simplicity.  Here, $\mu$ is merely a label to distinguish
different weightings; the $\tilde C^\mu$ are a linear combination of the old power spectra with weight $w^\mu_i$. Note that if a weighting satisfies,
\be
\label{eqn:cmbfree}
\sum_i w^\mu_i = 0
\ee
it is not sensitive to the CMB.

We would first like to find the CMB weighting $w^{\rm CMB}$ which will be statistically orthogonal to the $N-1$ linear combinations which satisfy the CMB--free condition (Eq. \ref{eqn:cmbfree}). 
We would also like this weighting to be properly normalized so that,

\be 
\label{eqn:cmbnorm}
\sum_i w^{\rm CMB}_i = 1 .
\ee

To satisfy the orthogonality condition it helps to work in a primed space defined by a linear
transformation via
\be
w'^\mu_\alpha = \sum_i L_{\alpha i} w^\mu_i
\ee
where $L$ is the Cholesky decomposition of the bandpower error
covariance matrix, $\Sigma = LL^T$.  The advantage of the primed
space is that the basis vectors in the primed space correspond to power spectra whose
errors are statistically orthogonal; i.e., with the weightings set so that
$w'^\mu_\alpha = \delta_{\mu \alpha}$ (now setting $\mu = 1...N$) the errors in the corresponding power spectra satisfy
\be
\langle \delta \tilde C^\mu \delta \tilde C^\nu \rangle = \delta_{\mu \nu}.
\ee

The primed weights that satisfy the CMB--free condition satisfy 
\be 
\sum_{\alpha, i} L^{-1}_{i\alpha} w'_\alpha = 0.
\ee
Thus any power spectrum with primed weighting proportional to 
\be
w'^{\rm CMB}_\alpha = \sum_i L^{-1}_{i \alpha}
\ee
is perpendicular to any vector satisfying the CMB--free condition, as
one can easily verify.  To find the CMB weighting in the unprimed
space we perform the inverse transform and normalize to satisfy the
normalization condition (Eq.~\ref{eqn:cmbnorm})
\bea
w^{\rm CMB}_k &=& \sum_{i,\alpha} L^{-1}_{k\alpha} L^{-1}_{i,\alpha} \left[\sum_{i,k,\alpha} L^{-1}_{k\alpha} L^{-1}_{i\alpha}\right]^{-1} \\
&=& \sum_{i} \Sigma^{-1}_{ik} \left[ \sum_{i,k} \Sigma^{-1}_{ik} \right]^{-1}.
\eea
Note that this is the expression for inverse-variance weighting.

Our remaining task is to construct the $N-1$ CMB--free weightings in a
manner that leaves them all statistically orthogonal to the CMB
weighting.  We do so by applying the Gramm-Schmidt procedure in the
primed space.  This gives us $N-1$ orthogonal vectors that are all
orthogonal to the CMB direction as well, that we will call
$v'^\mu_\alpha$ for $\mu = 2, N$.  The weightings in the unprimed
space are then given by 
\be
v^\mu_i = L^{-1}_{i \alpha} v'^\mu_\alpha.  
\ee

We now define the matrix $W$ so that
\bea
\label{eq:wcmb}
W_{1i} &=&  w^{\rm CMB}_i \nonumber \\
W_{\mu i} & = &  v^\mu_i \{ {\rm for } \  \mu = 2...N\}.
\eea
This matrix defines the linear combinations of the power spectra that
have all the properties we desire.  The first row is the optimal CMB
weighting and subsequent rows give the $N-1$ CMB--free linear
combinations.  All the linear combinations are statistically
orthogonal; i.e., the covariance matrix for the new power spectra,
$W^T\Sigma W$, is diagonal.  Furthermore, $W$ is non-singular so we have
not lost any information through this re-weighting.  

\subsection{Modeling the foreground residuals with principal
  components}
  
\label{sec:lincombopca}

With the weight matrix $W$ defined, we can constrain the foreground power in the $N-1$ CMB--free power spectra by running an MCMC chain. Despite the large number of parameters and power spectra, this analysis is fast in practice because the foreground model consists of simple analytic forms and precomputed templates, and does not depend on any costly Einstein-Boltzmann solver or lensing models. For the set of foreground parameters at each step in this chain, we calculate the corresponding foreground contribution to the CMB linear combination. These $\ell$-dependent contributions form the columns of the $Y$ matrix in a principal component analysis (see Appendix~\ref{app:pca}). Following the PCA procedure, we have a few principal components and priors on their amplitudes which must be marginalized over in a separate chain which uses only the CMB linear combination. 

\subsection{Discussion of linear combination analysis}

The weightings which make up the CMB and CMB--free linear combinations depend on the bandpower covariance matrix, and thus on the noise properties of the instrument, any filtering which is performed, and on the true power spectrum on the sky. The principal components for the foreground residuals also depend on the choice of foreground model. For a \Planck temperature-only forecast and for our fiducial model, we present the results of a linear combination analysis. 

In the top panel of \fig \ref{fig:lincombo} we plot the weights for
the CMB linear combination as a function of $\ell$. At high $\ell$
where the measurement is noise dominated, nearly all of the CMB
information is contained in the 217\,GHz map which is the least
noisy. At lower $\ell$ where we become
dominated by cosmic variance, the CMB information comes from the
channels with the lowest foreground contamination.

Given these weights, we plot in the middle panel of \fig \ref{fig:lincombo} the foreground contribution to the CMB linear combination and the error bars on this new power spectrum. Also shown are the error bars for the 217\,GHz channel alone for comparison. The maximum improvement is at $\ell=2000$ where the error bars tighten by a factor of 1.4. We also see that the dominant contribution to the foreground power in the $\ell$ range where \Planck is most sensitive to the CMB is the radio Poisson, followed by the DSFG clustering.

Finally, we perform a PCA on the foreground residuals in the CMB
linear combination. The first several principal components are shown
in the bottom panel of \fig \ref{fig:lincombo}. We find that all of
the variation $>1$\,\muK2 can be described by two amplitude parameters,
as compared to the 14 parameters which govern these foregrounds. Another way to put this is
that using the CMB--free combinations we can clean out almost 140\,\muK2 of foregrounds (at 
$\ell=3000$), leaving only tens of \muK2 of residual uncertainty, modeled with the two 
principal components.

\section{Results}
\label{sec:res}

With the model and forecasting tools in place, we are ready to present
the results of our main analysis.  We want to find which components
can potentially cause large biases in an analysis of \Planck data, so
that we can model them with sufficient care.  We would also like to
know how much constraining power is reduced due to foreground
confusion.  Could significant improvements in cosmological parameter
constraints be achieved by using additional data and/or modeling?  To
answer these questions, we run a suite of forecasting analyses aimed
at singling out the effects of each foreground contribution.

The next sub-sections are organized as follows. First we examine the importance of each component by turning it on or off in the analysis. For the components which we deem important, we check whether our modeling is sufficient to protect the cosmological parameters from biases, both at the \Planck and \Planck+\Ground sensitivity levels. We then examine the degradation in statistical errors from the need to marginalize over the foregrounds, and finally we explore the impact of adding in ground-based data.

\subsection{Importance of the Different Components}
\label{sec:importance}

\begin{figure}
\centering
\includegraphics[width=3in]{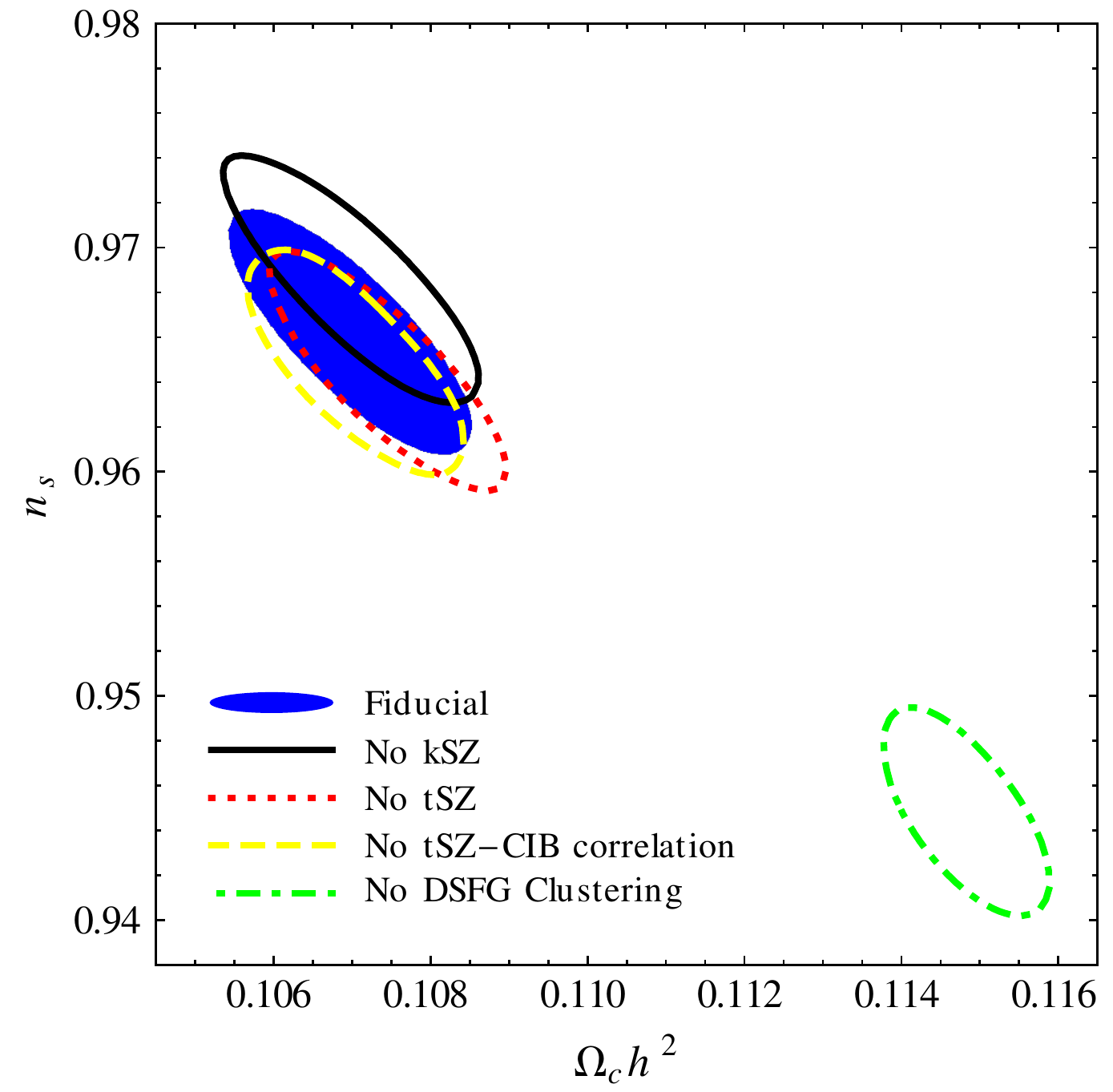}
\vspace{5mm}
	
\includegraphics[width=3in]{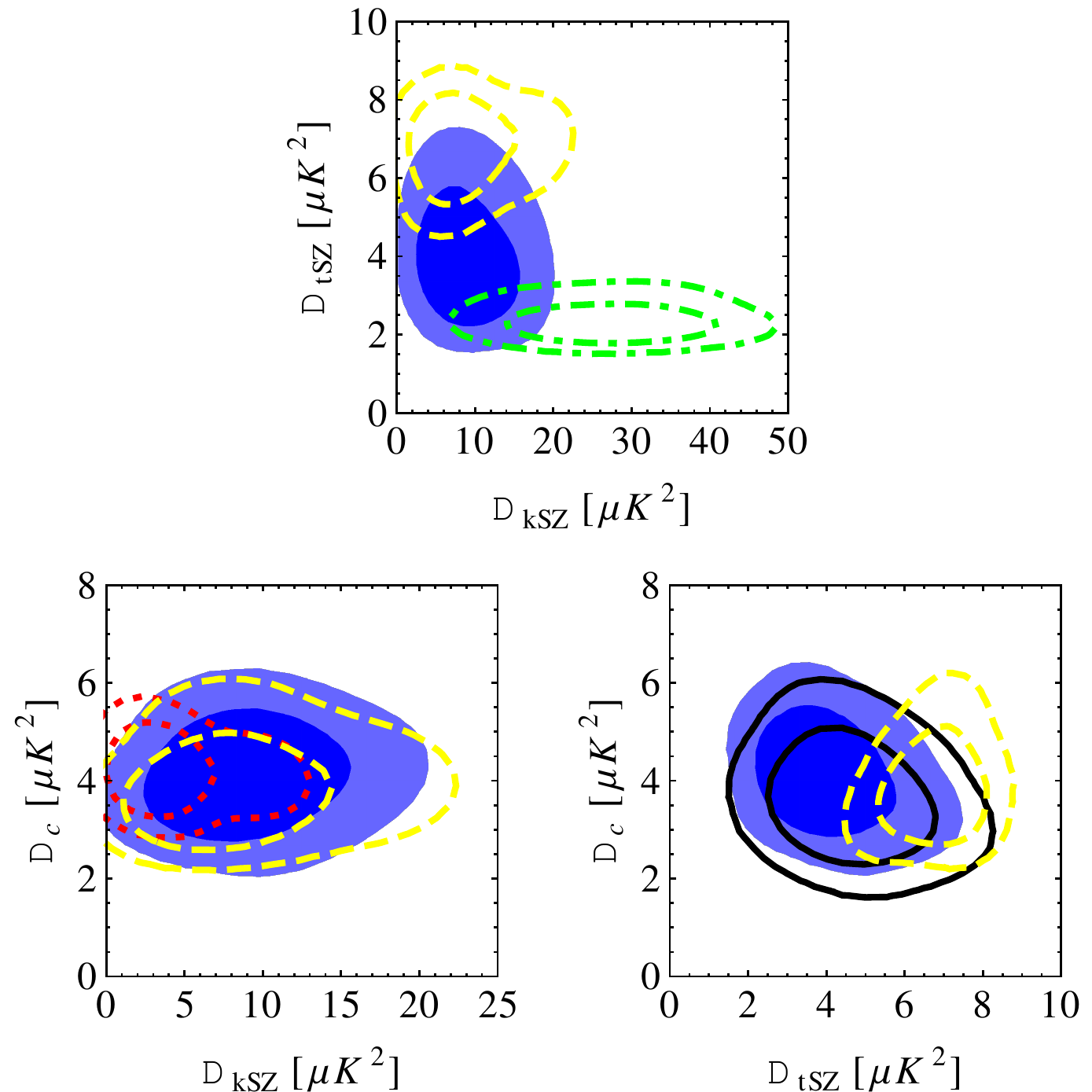}
\vspace{5mm}

\begin{tabular}{l|c|c|c|c}
\hline\hline
 & Data & Analysis & $\Delta\chi^2$ & $N_\ell$ \\ \hline
No tSZ & $\mathcal{D}_{tSZ} = 4.3$ & $\mathcal{D}_{tSZ} = 0$ & 43 & 6 \\ 
No kSZ & $\mathcal{D}_{kSZ} = 4.2$ & $\mathcal{D}_{kSZ} = 0$ & 2 & 1\\ 
No DSFG Clustering & $\mathcal{D}_{C} = 3.9$ & $\mathcal{D}_{C} = 0$ & 791 & 1930\\ 
No tSZ-DSFG Corr. & $r_{tSZ,C} = -.3$ & $r_{tSZ,C} = 0$ & 3 & 1
\end{tabular}
\caption{68\% (and 95\% in the bottom panel) confidence contours for a suite of test cases examining the effect of neglecting to model different foregrounds. Unless explicitly stated above, other parameters were included in the data at their fiducial values listed in \tab \ref{tab:params} and were marginalized over in the analysis. $N_\ell$ corresponds to the maximum number of $\ell$-bins per power spectrum one could use and still detect the error in modeling at 3 sigma (see Sec.~\ref{sec:importance} for further discussion).
}
\label{fig:planckbias}
\end{figure}

\fig \ref{fig:planckbias} shows the effect of removing four foreground components---the DSFG clustering, tSZ, kSZ, and tSZ-DSFG correlation---one at a time from the analysis model, while they are actually present in the simulated data at their fiducial value. We present the results by plotting likelihood contours in the $n_s$ and $\Omega_c h^2$ plane, since changes in those two parameters affect the primary CMB at the smallest scales and are the most susceptible to foreground biases. We also show the amplitudes of the clustering and
SZ effects as their $\ell$-space shapes make them most degenerate with cosmological parameters. All of the chains in this section include only \Planck power spectra in the simulated data.

We expect the DFSG clustering to be extremely important to model since it is the
second largest foreground contribution to the CMB linear combination
in the $\ell$-range where \Planck is most sensitive. When marginalized over, this contribution is constrained to be $10.5\pm0.6$\,\muK2 at $\ell=1500$, so setting it to zero is about an 18$\sigma$ change. The dot-dashed 
green contours in \fig \ref{fig:planckbias} show that this is compensated by a 
systematic bias of $7\sigma$ is $n_s$ and $11\sigma$ in $\Omega_c h^2$, along with
an increased kSZ power to about 30\,\muK2. Using the 
middle panel of \fig \ref{fig:lincombo} as a visual guide, we can examine how this 
happens. Though the kSZ increases to
compensate for the missing clustering power at high-$\ell$, its shape is
flatter than the DSFG contribution to the CMB linear combination, so $n_s$ 
decreases to roughly remove the extra power at a low-$\ell$.

At 217\,Ghz, the kSZ power in our fiducial model is a factor of 20 times smaller 
than the DSFG clustering and is therefore (in terms of cosmological parameter 
estimation) less troublesome. However, due to its identical frequency dependence 
to the CMB, we do expect the amplitude of the kSZ signal to be degenerate with 
cosmological parameters. The solid black contours in \fig \ref{fig:planckbias} 
demonstrate the bias introduced if the kSZ component is omitted from the 
foreground modeling. We find a roughly 0.5$\sigma$ bias in $n_s$ as it increases 
to try to fill in the missing 4.2\,\muK2 of kSZ power. The DSFG and tSZ plane 
(bottom right panel of \fig \ref{fig:planckbias}) shows that the other foregrounds 
are largely unaffected.

The thermal SZ component is neither frequency independent, nor does it
contribute as much power as the DSFG clustering, so we do not expect a
bias as large as in either previous case. It does, however, project
into the CMB linear combination to an $\ell$-shape very similar to the
CMB itself (see the middle panel of \fig \ref{fig:lincombo}), making
it more likely to be degenerate with cosmological parameters. From the
results, we see about 0.3$\sigma$ biases in each of $n_s$, $\Omega_c
h^2$, and $\Omega_b h^2$.

Finally, we consider neglecting a 30\% tSZ-DSFG correlation, a
value on the high end of expected correlation, but still consistent
with \citet{shirokoff10}. We expect this to have the smallest effect on
the cosmological parameters since the power contribution is
sub-dominant to all of the other foreground components at all
frequencies which appear in the CMB linear combination at
$>1\%$. While the measured tSZ amplitude is biased at a few sigma as
it raises to compensate for the missing power, the effect is not large
enough to significantly impact any of the cosmological parameters.

One question is whether any of these analysis errors would be caught
by a goodness-of-fit test.  To address this question we present
$\Delta \chi^2$ values in the table in Fig.~\ref{fig:modelerrors}. 
We can expect rms fluctuations in $\chi^2$ to be $\sqrt{N_b}$ where
$N_b$ is the total number of bandpowers which is roughly equal to the
number of degrees of freedom.  If one is searching for signs of a contaminant
that is very slowly varying in $\ell$, then one would bin coarsely to
reduce the statistical fluctuations in $\chi^2$, to make a more
stringent goodness-of-fit test.  

We define $N_\ell$ to be the number of $\ell$-space bins such that the
absolute $\Delta\chi^2$ from the fit is 99.7\% inconsistent with
random fluctuations.  Thus we have $N_\ell = (\Delta\chi^2)^2/(9
\times N_{\rm spec})$ where $N_{\rm spec}$ is the number of power
spectra (36 here).  We see that binning would not have to be coarse at
all to detect the poor fit caused by neglecting clustering.  We also
see that for the other entries in the table, binning would have to be
extremely coarse for the fits to be noticeably poor.  Indeed, the
binning would have to be coarser than is practical since the signals
of interest, as well as the contaminants, would vary significantly
across a bin.  We conclude that only the ``no clustering'' case would
produce a noticeably bad fit for \Planck only.

\subsection{Modeling Sufficiency}
\label{sec:model_sufficiency}

\begin{figure}
\centering
\includegraphics[width=3in]{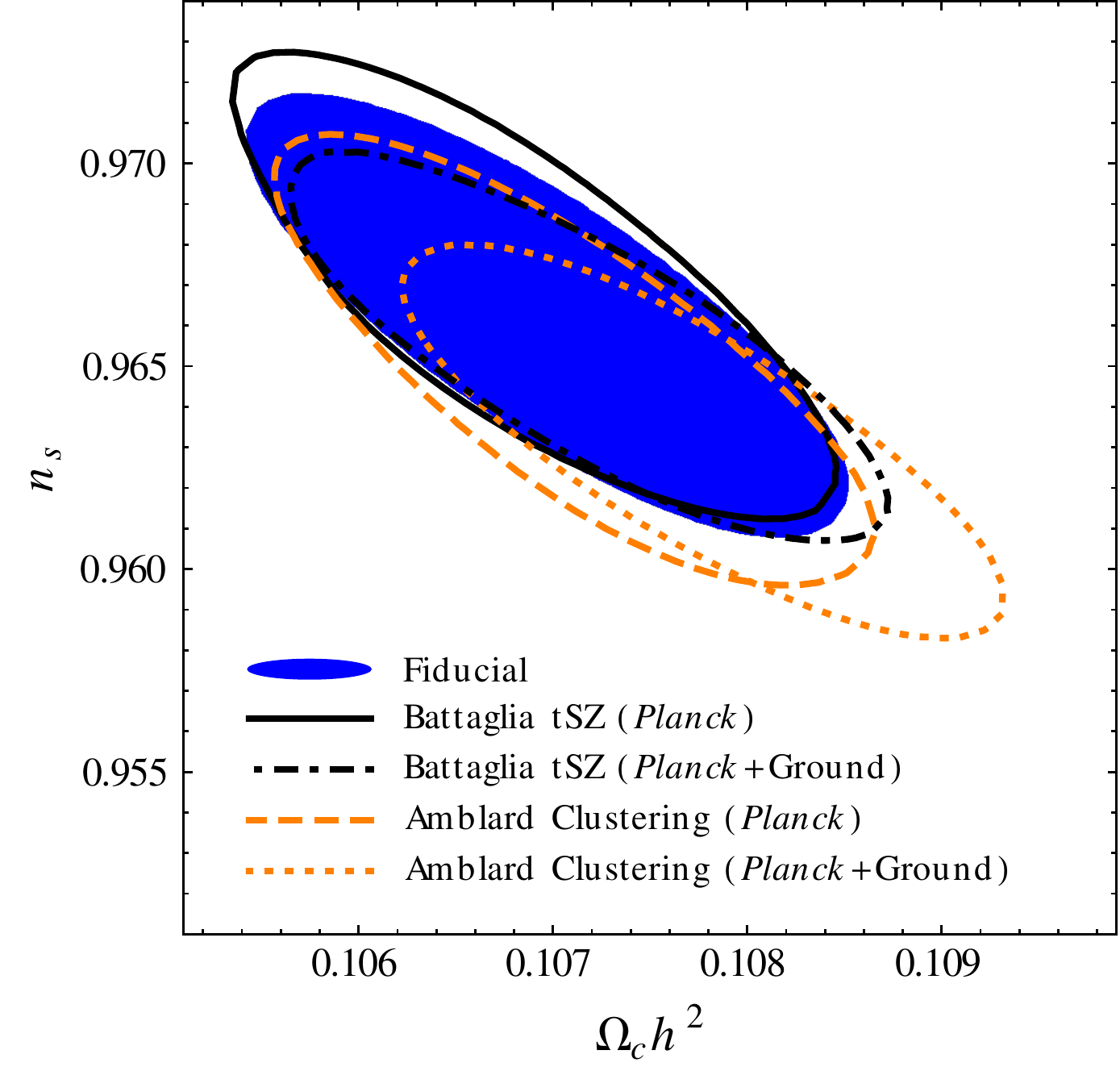}
\caption{The effect on cosmological parameters from trying to fit
  our model to simulated data which includes (orange) the \citet{battaglia10} tSZ
  template and (black) the \citet{amblard07} clustering template. 
  These two models are the most dissimilar to ours, and thus show our 
  model can protect against biases of a few percent up to \Planck sensitivity.
  The inclusion of \Ground data necessitates more detailed modeling
  of only the clustering.}
\label{fig:modelerrors}
\end{figure}

Given the demonstrated importance of the foreground components, we
would now like to see if our modeling is sufficient to protect the
cosmological parameters from biases if we have modeled the
components, but modeled them {\em incorrectly.} In this section we
consider the DSFG clustering and the tSZ effect.

For the DSFG clustering, we turn to the models plotted in \fig
\ref{fig:cibclustering}. Our parameterization should have the most
trouble reproducing the \citet{amblard07} model, which switches to a
power-law (as a consequence of non-linear clustering) at $\ell\approx 2500$
rather than at $\ell = 1500$ as in our fiducial model. The orange
contours in \fig \ref{fig:modelerrors} show the results obtained when
fitting our model to simulated power spectra that assume the
\citet{amblard07} clustering template. As we had hoped, for the case of
\Planck only (solid lines), there is no significant biasing.

We also explore the ability of our tSZ principal component model
\citep[based on the analytic model of][]{shaw10} to encompass the
variations in the tSZ models shown in \fig
\ref{fig:tsz_pca_models}. We elect the \citet{battaglia10} one as the most
dissimilar, since it lacks the effects of radiative cooling, and
should be the most difficult for the \citet{shaw10} model to
reproduce.  Despite these differences, \fig \ref{fig:modelerrors}
shows that for \Planck the model is sufficient to encompass the shape
uncertainty and protect cosmological parameters.

When we add in \Ground (dashed lines), the requirements on the
modeling accuracy are more stringent.  For the clustering case, we see
an almost $1\sigma$ bias from using our fiducial model when the true
model is the \citet{amblard07} clustering template. Analyses with
current data can tolerate much more discrepant clustering shapes
\citep{dunkley10}.  For future \Planck+\Ground analyses, the
clustering shape will need to be modeled more accurately.  For tSZ the
modeling appears to be more robust; tSZ-induced biases are small even
in the \Planck+\Ground case.

\subsection{Statistical Error Increase with and without Auxiliary Data}

\begin{table*}
\centering
\caption{Statistical Error Degradation}
\begin{tabular}{l|c|c|c|c|c|c|c|c|c|c}
\hline\hline
 & $\Omega _bh^2$ & $\Omega _ch^2$ & $\Theta$  & $n_s$ & $\ln(10^{10} A_s)$ & $\mathcal{D}_D$ & $\mathcal{D}_R$ & $\mathcal{D}_C$ & $\mathcal{D}_{tSZ}$ & $\mathcal{D}_{kSZ}$ \\ \hline
 \Planck (fgs fixed) & $\mathbf{1.1\!\times\!10^{-4}}$ & $\mathbf{1.0\!\times\!10^{-3}}$ & $\mathbf{2.6\!\times\!10^{-4}}$ & $\mathbf{3.0\!\times\!10^{-3}}$ & $\mathbf{1.3\!\times\! 10^{-2}}$ & -- & -- & -- & -- & --\\ 
 \Planck (fgs marginalized) & 110\% & 100\% & 100\% & 120\% & 110\% & \textbf{3.4} & \textbf{6.0} & \textbf{1.3} & \textbf{1.0} & \textbf{4.4} \\
 \Planck+\Ground & 100\% & 100\% & 100\% & 110\% & 100\% & 10\% & 50\% & 40\% & 60\% & 60\% \\
 \Planck (Clean DSFG) & 100\% & 100\% & 100\% & 110\% & 100\% & 60\% & 100\% & & 100\% & 100\% 
\label{tab:staterrors}
\end{tabular}
\tablecomments{Entries in bold show the standard deviation for each parameter for 
that particular row. Subsequent percentages reference the bold entries in the same 
column. Dashes indicate the parameter was fixed, while blanks mean the parameter 
is not applicable to that case. The normalization parameters $\mathcal{D}_x$ are in units of \muK2. The different cases correspond to: (fgs 
fixed) Fixing all of the 
foregrounds at their fiducial values. (fgs marginalized) Marginalization over our 
full foreground model. (+Ground) Also including \Ground auto and cross spectra in the simulated data. (Clean DSFG) Assuming 90\% reduced clustering power due to cleaning from higher frequencies.}
\end{table*}

We have demonstrated the possibility of $\sigma$-level biases in
cosmological parameters arising from failure to model foregrounds. To
prevent these biases, the foregrounds need to be jointly estimated or
marginalized over.  We now turn to two questions: 1) How much do the
cosmological parameter statistical errors degrade due to foreground
uncertainty? and 2) How much can be gained from using other data to
constrain foregrounds and thereby reduce that degradation?

The top two rows of \tab \ref{tab:staterrors} show the effect of marginalizing
over our entire foreground model as opposed to fixing it at fiducial values. 
In each row, the difference from 100\% is the
percent degradation due to foreground
marginalization. The second row shows the degradation is limited to 20\% for $n_s$ and 10\% for $A_s$ and $\Omega_b h^2$. We see
no degradation in $\tau$ and $r$ since they are mainly constrained by
large scales where the extragalactic foregrounds we consider are
negligible. The dark energy equation of state $w$ is unaffected
because it is mainly constrained by our $\pm 0.3$ prior.

Ground data can help reduce this degradation by better constraining the 
foregrounds using auto and cross spectra that are more sensitive at small scales. 
The improvement from adding these to the simulated data is shown in
the row labeled \Planck+\Ground.  The measurement of 
DSFG shot noise is improved ten-fold, with the clustering and SZ effects also 
tightened by a factor of two. The radio amplitude is improved through constraints 
on the spectral dependence, and could be further improved though a prior on 
$\gamma_R$ from \Ground source counts. The 
effect on the cosmological parameters is to remove essentially
all of the degradation
we incurred from marginalizing over the foreground model. 

Above about 300\,GHz, the DSFGs are the dominant source of 
anisotropy power on all scales.
Correlations with maps at these higher frequencies, for
example maps from \Planck bands above 
217\,GHz or \textit{Herschel}, 
can be used to place tight constraints on the DSFG components, at the price of
requiring more sophisticated modeling for the spectral dependence and shape.
Even with such modeling, the correlations are no longer fully coherent
across frequencies so there is a limit to how much of the DSFG power can be 
``cleaned out" of the lower frequency maps. Following 
results in \citet{knox01}, which assumes a redshift dependent grey-body emissivity 
density tracing the linear matter power spectrum, we assume that we could clean 
out 90\% of DSFG clustering power at the lower frequencies. As in the previous 
case of adding in \Ground data, this again is enough to eliminate nearly all of 
the degradation on cosmological parameters.

\section{Conclusions}
\label{sec:disc}

To make full use of \Planck's very small statistical error on CMB
power spectra out to $\ell \sim 2500$, without introducing significant
bias in the cosmological parameters, we must include contributions
from extragalactic foregrounds and secondaries in our model of the
data.  Here we have presented a model of these contaminants, based on
the latest data and modeling developments, and demonstrated its ability to
remove biases in an 8-parameter cosmological model.  The foreground
model has 17 parameters -- many more than any extragalactic foreground
model used in analysis of CMB data to date.  Despite the large number
of nuisance parameters, marginalizing over all of them only increases
statistical uncertainties in the cosmological parameters by, at most
10 to 20\%.  Almost all of this degradation can be avoided by
inclusion of ground-based data or higher frequency \Planck bands.

Our model includes Poisson components from both radio galaxies and
DSFGs, a clustering component due to DSFGs, contributions to kSZ power
from patchy reionization, as well as after reionization is complete,
and tSZ power.  If kSZ power and tSZ power are at our fiducial values
(slightly higher than the preferred values given current
high-resolution ground-based data) then ignoring them in an analysis
of \Planck data would produce small, almost negligible biases, to
cosmological parameter estimates.  On the other hand, ignoring the
clustering of DSFGs, would lead to a very large bias in cosmological
parameters.

To avoid having to marginalize over these 17 parameters every time a
new cosmological model is analyzed, we broke our procedure up into a
two-step process, with the first step independent of the model of the
primary CMB power spectra.  The second step is an analysis of the CMB
power spectra estimated in the first step, with a small number of
foreground template amplitude parameters to marginalize over.  The
shapes of these templates, and priors on their amplitudes, are also
outputs of the first step.  Only the second step needs to be repeated
in order to get constraints on the parameters of a new model of the
primary CMB power spectra.  

Looking toward the near future, the model will definitely evolve, increasing the faithfulness with which
it represents reality, as we gain more information from the
CMB-dominated channels in \Planck, higher-frequency \Planck channels, 
higher-resolution ground-based data (SPT, SPTpol and ACTpol) and
higher-resolution, higher-frequency space-based data ({\it Herschel}).
One could easily use our foreground model to study potential biases in extensions of the
primary cosmological model, to include, for example, departure of the
Helium mass fraction from its BBN-expected value, or the number of
effective neutrino species from its standard model value.

\begin{acknowledgements}
We benefited from conversations with A. Challinor, J. Dunkley,
 G. Efstathiout, F. Finelli, S. Gratton, W. Holzapfel, C. Reichardt,
 D. Scott, and G. de Zotti. Part of the research described in this
 paper was carried out at the Jet Propulsion Laboratory, California
 Institute of Technology, under a contract with the National
Aeronautics and Space Administration. L. Shaw acknowledges the support
of Yale University and NSF grant AST-1009811. L. Knox and M. Millea
 acknowledge support from NSF grant 0709498.
\\
\end{acknowledgements}

\bibliographystyle{apj}
\bibliography{planck_spt}

\begin{appendix}

\section{A. CMB Linear Combination Generalization to Off-Diagonal Correlations}
\label{app:lincombo}
The method for constructing a best estimate of the CMB presented in
Section \ref{sec:lincombo} assumes only temperature power spectra, and
a covariance which is diagonal in $\ell$. The generalization to
include polarization and mode-mode coupling induced by sky masking is
presented here. The math is, infact, indentical for the two scenarios,
so in this appendix we'll refer to polarization types with the
understanding that we could just as well be talking about different
values of $\ell$.

The added difficulty in dealing with different power spectrum types
(for simplicity here just TT and EE) comes from the fact that we
cannot arbitrarily create linear combinations which sum them. For
example, $C' = C^{TT}_{100\,\mathrm{GHz}} -
C^{EE}_{100\,\mathrm{GHz}}$ neither preserves CMB normalization, nor
can we be sure it is CMB--free independent of model. To remedy this, we
make sure that in our construction, any linear combination we consider
must have the CMB signal cancel out for all but one type. For example,
$C' = C^{TT}_{100\,\mathrm{GHz}} - C^{EE}_{100\,\mathrm{GHz}} +
C^{EE}_{143\,\mathrm{GHz}}$ is a valid linear combination.

We start by considering the covariance matrix for the TT and EE spectra.

\begin{equation}
\left[
\begin{array}{cc}
\Sigma_{TT} & \cdots \\
\cdots & \Sigma_{EE}
\end{array}
\right]
\end{equation}
By creating the single-type weight matrix (\eq \ref{eq:wcmb}) for each of the diagonal blocks, we can cancel the CMB out of all but two weightings. The new covariance will look like,
\begin{equation}
\left[
\begin{array}{ccc}
W^{T}_{TT} & 0 \\
0 & W^{T}_{EE}
\end{array}
\right]
\left[
\begin{array}{cc}
\Sigma_{TT} & \cdots \\
\cdots & \Sigma_{EE}
\end{array}
\right]
\left[
\begin{array}{cc}
W_{TT} & 0 \\
0 & W_{EE}
\end{array}
\right]
=
\left[
\begin{array}{cc}
\left(
\begin{array}{cc}
\sigma_{TT} & 0 \\
0 & \ddots
\end{array}
\right)
& \cdots \\
\cdots &
\left(
\begin{array}{cc}
\sigma_{EE} & 0 \\
0 & \ddots
\end{array}
\right)
\end{array}
\right]
\end{equation}
Under a permutation to place the two CMB weightings at the front, the covariance becomes,
\begin{equation}
\left[
\begin{array}{cc}
\left(
\begin{array}{cc}
\sigma_{TT} & \cdots \\
\cdots & \sigma_{EE}
\end{array}
\right)
& \cdots \\
\cdots & \ddots
\end{array}
\right]
\equiv
\left[
\begin{array}{cc}
\Sigma_{cmb} & \Sigma_{cross}^T \\ \Sigma_{cross} & \Sigma_{diff}
\end{array}
\right]
\end{equation}
where we've labeled $\Sigma_{cmb}$ as the covariance between TT and EE estimates, $\Sigma_{diff}$ as the covariance of the CMB--free differenced spectra, and $\Sigma_{cross}$ as the cross-correlation between the two. We now would like to do one final reweighting in an attempt to zero out the cross-correlation. The reweighting should leave the differenced spectra unchanged, should not add TT and EE together, but will add CMB and CMB--free power spectra. Note that this will continue to satisfy our earlier condition that all but one CMB type canceling out. The reweighting matrix will look like,
\begin{equation}
\left[
\begin{array}{cc}
I & 0 \\ W' & I
\end{array}
\right]
\end{equation}
The new covariance must satisfy,
\begin{equation}
\left[
\begin{array}{cc}
I & W'^{T} \\ 0 & I
\end{array}
\right]
\left[
\begin{array}{cc}
\Sigma_{cmb} & \Sigma_{cross}^T \\ \Sigma_{cross} & \Sigma_{diff}
\end{array}
\right]
\left[
\begin{array}{cc}
I & 0 \\ W' & I
\end{array}
\right]
=
\left[
\begin{array}{cc}
\Sigma'_{cmb} & 0 \\ 0 & \Sigma_{diff}
\end{array}
\right]
\end{equation}
Solving for $W'$ yeilds,
\begin{equation}
W' = - \Sigma_{diff}^{-1} \Sigma_{cross}
\end{equation}

\section{B. Principal Component Analysis for Power Spectra}
\label{app:pca}
In this paper we use a principal component analysis (PCA) to reduce the dimensionality of the tSZ astrophysical parameter space (Sec. \ref{sec:tsz}) and of the entire foreground contribution to the CMB linear combination (Sec. \ref{sec:lincombopca}). Here we present in more detail the procedure used in those sections.

Given $n_{sim}$ realizations of an $n_\ell$-length power spectrum, drawn from a statistically significant sample of parameter space, we first form the $[n_{\ell}\times n_{sim}]$ matrix $Y$. In each column of $Y$ we place the deviation from the mean power spectrum for that realization. This matrix is then subject to a singular value decomposition, 
\begin{equation}
\label{eq:svd}
Y = U S V^T \;,
\end{equation}
where the columns of $U$ contain the orthogonal basis vectors, $S$ is
a diagonal matrix of the singular values and the columns of $V$ are
the principle component weights. The $i$-th realization can be written as
\begin{equation}
\label{eq:pcasum}
y^{(i)}_\ell = \sum_{\mu} \Phi_\ell^{\mu} w^{(i)}_{\mu}\;,
\end{equation}
where the singular value-weighted orthogonal basis vectors are
\begin{equation}
\Phi^\mu_\ell  = \frac{1}{\sqrt{n_{sim}}} U_{\ell\mu} S_{\mu\mu} \;.
\end{equation}
and the $w_\mu$ are the weights,
\begin{equation}
w^{(i)}_\mu = \sqrt{n_{sim}} \, V_{i\mu} \;.
\end{equation}
Because the singular values are in decreasing order, we can truncate the sum in Eq. \ref{eq:pcasum} at some small value of $\mu$ and still accurately describe each realization. Furthermore, the distribution of weights $P(w_\mu)$ sampled over $n_{sim}$ realizations provides a prior on our principal component amplitudes equivalent to the parameter space which was sampled to produce the $Y$ matrix.

\end{appendix}

\end{document}